\documentclass{egpubl}
\usepackage{pg2026s}

\WsConferencePaper

\CGFccby

\usepackage[T1]{fontenc}
\usepackage{dfadobe}  

\usepackage{cite}  
\BibtexOrBiblatex
\electronicVersion
\PrintedOrElectronic
\ifpdf \usepackage[pdftex]{graphicx} \pdfcompresslevel=9
\else \usepackage[dvips]{graphicx} \fi

\usepackage{egweblnk} 

\usepackage{url}
\usepackage{amsmath,amssymb,graphicx}
\usepackage{xcolor}
\usepackage{lipsum} 

\usepackage[switch]{lineno}
\usepackage{booktabs}
\usepackage{algorithm}
\usepackage{algpseudocode}
\usepackage{algorithmicx}
\usepackage{xcolor}

\title[Volumetric Simulation and Rendering of Heat-induced Optical Turbulence at Interactive Rates]%
      {Volumetric Simulation and Rendering of Heat-induced Optical Turbulence at Interactive Rates}

\author[W. Yuan et al.]
{\parbox{\textwidth}{\centering Wanqi Yuan$^{1}$,
        Ethan Chung$^{2}$,
        Man Luo$^{2}$,
        Suren Jayasuriya$^{3}$\orcid{0000-0001-7143-4429},
        Huaijin Chen$^{2}$\orcid{0000-0002-7337-149X},
        Jinwei Ye$^{4}$\orcid{0000-0001-7780-7943},
        and Nianyi Li$^{1}$\orcid{0000-0002-4172-4940}
        }
        \\
{\parbox{\textwidth}{\centering $^1$Clemson University, $^2$University of Hawai\textquoteleft i at M\=anoa, $^3$Arizona State University, $^4$George Mason University
       }
}
}

\begin{document}

\teaser{\vspace{-1cm}
 \includegraphics[width=0.9\linewidth]{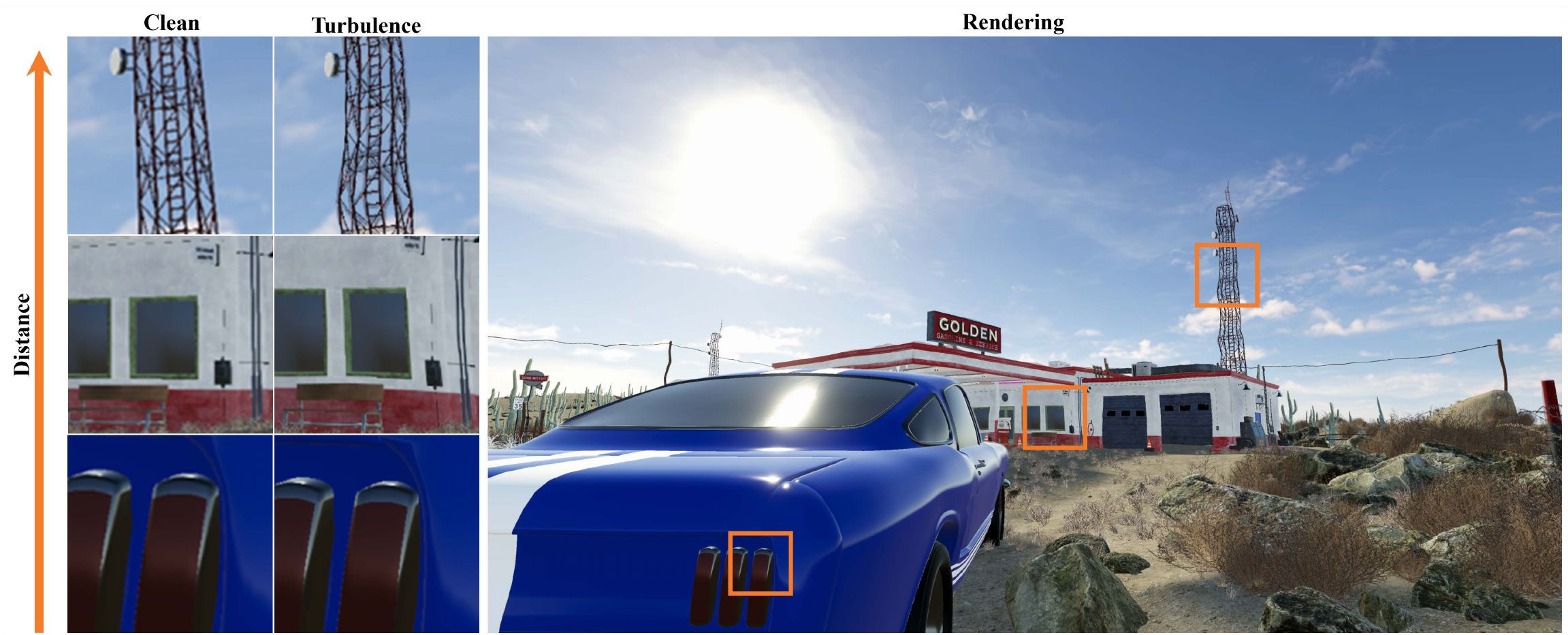}
 \centering
  \caption{A 3D refractive volume rendered by our method. Left: clean and turbulence-affected crops at three depths (orange boxes, near $\rightarrow$ far),
showing distortion growing with distance.
Right: the full rendered frame.}
\label{fig:teaser}
}

\maketitle
\begin{abstract}
Heat-induced optical turbulence produces complex, depth-dependent image distortions that are challenging to simulate interactively because thermally-driven flow must be coupled with refractive light transport. Existing methods either rely on single-view 2D screen-space warping that breaks multi-view coherence, or employ high-fidelity offline solvers that preclude interactive use. We present an interactive, fully 3D Lagrangian framework that models the complete simulation pipeline from thermal transport to density variation to optical refraction.
Our system augments Smoothed Particle Hydrodynamics (SPH) with temperature transport, buoyancy, and a convective perturbation term to capture rising plumes and turbulent mixing. We render the resulting continuous refractive-index field via curved ray tracing to model light bending in 3D. To reconcile physical fidelity with interactive performance, we introduce \textit{curvature-adaptive step-size} integration for curved-ray tracing, bounding the per-step angular deviation to concentrate samples near strong refractive-index gradients while relaxing them in smooth regions. The system runs at interactive rates (about 40\,fps in our prototype) and matches depth-dependent, multi-view-consistent distortions observed in real video captures more closely than image-based baselines. 


\end{abstract}  
\section{Introduction}


Heat-induced optical turbulence is an omnipresent source of image degradation in long-range imaging, autonomous driving, and outdoor multi-camera systems. The visible shimmer arises from a coupled physical process: heat transport drives buoyant flow and density variation, inducing a spatially varying refractive index that bends light along curved paths. Faithfully reproducing this phenomenon is essential for generating controlled training data for turbulence-robust vision algorithms~\cite{mao2021accelerating,chimitt2022,saha2024turb,zhang2024spatio}, for evaluating multi-view stereo and depth estimation under atmospheric distortion, and for immersive rendering in VR/AR with physically grounded environmental effects~\cite{Pfaff2010,bhattarai2020embedded,lanman2013near}. A central challenge is that existing simulators either operate in 2D screen space---breaking multi-view coherence---or are too expensive for interactive use.

Most existing approaches leave a key gap: they do not provide a shared, time-varying \emph{3D} refractive volume that multiple cameras can observe consistently.
High-fidelity offline methods based on Eulerian solvers, including Lattice Boltzmann variants~\cite{teixeira1998incorporating,zhao2006visual,jahanshaloo2013review}, can model thermally driven airflow and refraction with high accuracy, but their dense grid computations are typically too expensive for interactive use and large volumes.
Conversely, 
many interactive-rate methods are designed primarily as \emph{image-space distortion generators}~\cite{chimitt2020simulating,chimitt2022,saha2024turb}: 
they apply single-view screen-space warps to mimic turbulence but do not simulate a 3D medium and therefore cannot guarantee cross-view coherence or depth-dependent behavior. Because these methods produce independent distortions per view, they are fundamentally incompatible with multi-camera stereo, depth estimation, or any task that assumes geometric consistency across viewpoints, which is a critical requirement in computational photography and visual computing.

This paper targets a practical middle ground between these extremes.
Our goal is not to replicate every fine-scale aspect of fully resolved fluid turbulence, but to produce \emph{physically grounded and controllable} heat-haze behavior from an explicit heat $\rightarrow$ flow $\rightarrow$ density $\rightarrow$ refraction pipeline, efficient enough for interactive simulation and rendering. To this end, we present a unified model that couples a compressible Lagrangian SPH-based thermofluid model with volumetric refractive rendering.
Temperature evolves through heat exchange; buoyancy and pressure forces drive rising plumes and turbulent mixing; particle density is then mapped to a refractive-index field. Crucially, all cameras render through the \textit{same} dynamic 3D refractive volume using curved ray integration, so multi-view coherence is enforced by construction rather than approximated in screen space. To further reconcile fidelity with interactive performance, we introduce a \emph{curvature-adaptive step-size strategy for curved-ray integration} that bounds the per-step angular deviation, concentrating computation near strong refractive-index gradients and taking larger steps in smooth regions.

Additionally, we benchmark against \textit{real-world captures} exhibiting heat-turbulence effects and compare to representative image-based turbulence simulators, focusing specifically on depth-varying distortions and multi-view 3D consistency. 
To our knowledge, this is the first system that couples volumetric thermal-fluid simulation with curved-ray rendering to produce \emph{multi-view-consistent} heat-haze effects at interactive frame rates. Specifically, 
our major contributions are:

\begin{itemize} 
	\setlength\itemsep{0em}




    \item 
    We integrate heat-driven thermofluid simulation and volumetric refractive rendering into a single system that produces a shared 3D refractive volume observed consistently from multiple viewpoints. 
    An example is shown in Fig.~\ref{fig:teaser}.

    \item 
    We augment compressible SPH with temperature transport and buoyancy/pressure-driven motion to generate physically plausible plume dynamics and density variations suitable for interactive-rate refractive rendering (Sections~\ref{sec:thermalsph} and \ref{sec:sphsolver}).

    \item 
    We accelerate curved-ray integration by adapting step size to refractive-index variation, concentrating computation where distortions are visually significant and avoiding uniform oversampling elsewhere (Section~\ref{sec:adaptiveray}).

    \item 
    We evaluate against real-world video captures, image-based baselines and an offline curved ray tracer (Section~\ref{sec:results}), highlighting the benefits of a true 3D volumetric model for cross-view consistency and depth-varying distortion.
\end{itemize}

Our open-source code and scene files are publicly available at this URL: \url{https://github.com/WanqiYuan/TurbSimPG}.


\section{Related Work}
\label{sec:relatedwork}


\textbf{Thermal fluid simulation for atmospheric turbulence:} Simulating heat and haze effects from atmospheric turbulence at a fundamental level requires physics-based fluid solvers to compute spatially-varying changes in refractive index prior to rendering the final imagery. Eulerian methods, notably GPU-accelerated Lattice Boltzmann schemes coupled with heat transfer, can reproduce mirages and heat shimmer but require large memory footprints for high fidelity~\cite{teixeira1998incorporating,zhao2006visual,jahanshaloo2013review}. From the Lagrangian perspective, Smoothed Particle Hydrodynamics (SPH) was introduced in astrophysics~\cite{gingold1977smoothed,lucy1977numerical} and later adopted in graphics for interactive fluid simulation~\cite{desbrun,pbf,sph2024,ye2024monte,liusph2024}. Beyond early visually plausible effects, subsequent work explored more physically faithful turbulence modeling, often at higher computational cost~\cite{monaghan2002sph,violeau2007numerical,pbvfs,monaghan2011turbulence,adami2012simulating}. Monaghan~\cite{monaghan2002sph} developed compressible $\alpha$--turbulence models; Violeau~\cite{violeau2007numerical} analyzed multiple SPH turbulence schemes; and Adami et al.~\cite{adami2012simulating} studied detailed three-dimensional turbulence. In addition, researchers have also attempted to speed up SPH methods for modeling gaseous flows~\cite{bofastsph2016}. 

For modeling heat transfer for fluid simulation, SPH has been applied to conduction and convection in particulate and complex flows~\cite{CLEARY1999227,jeong2003smoothed,zhang2019finite}. More recent graphics work improves realism and robustness via vorticity-based detail enhancement~\cite{liu2021turbulent}, micropolar models for effects such as foam~\cite{bender2018turbulent}, and improved stability/convergence for both compressible and incompressible formulations~\cite{sissm}. While these methods advance SPH fidelity, they do not explicitly target the coupled dynamics characteristic of heat-induced air turbulence. We address this gap by extending~\cite{sissm} towards temperature-driven air turbulence suitable for downstream refractive rendering at interactive frame rates.

\textbf{Curved ray tracing in inhomogeneous medium:} After simulating the 3D volume's spatially-varying refractive index (induced from the underlying fluid properties), modeling light propagation through this medium demands curved ray tracing algorithms that extend beyond straight-line assumptions. Foundational work in this area includes Stam and Langu\'{e}nou~\cite{stam96} for ray tracing in non-constant medium, Berger et al.~\cite{berger1990ray} using Snell's law for mirages, Seron et al.~\cite{seron2005implementation} employing Fermat’s principle to implement accurate curved-ray tracing in offline renderers, and Gutierrez et al.~\cite{gutierrez2006simulation} solving the physically-based differential equation governing light trajectories in atmosphere. Recent graphics research~\cite{refractiveRTE,pediredla2020path} provides rigorous transport solutions in refractive medium including demonstrating differentiable rendering through continuously-varying index fields~\cite{Teh:2022:ANRT}. Fraboni et al.~\cite{fraboni2023} applied volume rendering techniques to the related problem of heat effects due to fire and explosions in graphics production systems. However, all these methods mentioned above are designed for offline rendering and not interactive applications. 

For faster rendering techniques, researchers have proposed Eikonal rendering for refractive objects~\cite{ihrke2007eikonal} and novel-view synthesis~\cite{eikonal2022}, although not applied to the particular problem of optical turbulence with varying refractive index changes over time. In the context of gradient-index optics, Yu et al.~\cite{yu2024adaptive} proposed an adaptive ray tracing algorithm for freeform medium utilizing index directional derivatives to minimize error. However, these optical engineering methods typically prioritize extreme numerical precision for static medium and rely on computationally heavy high-order solvers. In contrast, our system introduces a lightweight adaptive stepping scheme driven by local ray curvature. 
This approach allows us to resolve the high-frequency density gradients of dynamic SPH turbulence without the overhead of optical design solvers, enabling efficient interactive-rate performance.



\textbf{Image-based and procedural noise simulators for atmospheric turbulence:} In contrast to the fluid simulation and curved ray tracing presented above, researchers have proposed image-based simulators for fast generation of turbulence effects. These image-based simulators model optical turbulence through phase screens~\cite{schmidt2010numerical}, spatially varying point spread functions~\cite{bos2012technique,hardie2017simulation}, or phase-to-space transforms~\cite{mao2021accelerating,chimitt2022}. 
Such approaches can achieve interactive-rate generation; however they do not simulate a shared 3D refractive medium and therefore cannot guarantee valid depth effects or multi-view coherence. Recently, Saha et al.~\cite{daatsim} introduced depth-dependent effects for these image-based simulators using monocular depth estimation, but still without volumetric flow simulation. 

For non-physical simulators, 3D game engines often adopt procedural noise-based distortions~\cite{perlin85,lagae2010survey} which have been used for various atmospheric phenomena including turbulence~\cite{riley2021noise,saha2024turb}, clouds~\cite{webanck2018procedural}, and even volcanic eruptions~\cite{pretorius2024volcanic,lastic2022interactive}. These methods are both fast and volumetric in nature, but lack true physics correspondence to underlying flow or refractive-index dynamics. Our method aims for a practical middle ground: a simplified volumetric thermofluid model that enables depth-dependent distortions and multi-view consistency while quickly rendering plausible visual effects at interactive frame rates.



\section{A Unified Model for Interactive-Rate Heat-Induced Air Turbulence}

Our model calculates heat transfer between particles and sources, translating temperature changes into buoyancy and convective forces. These forces drive motion within a state-of-the-art Smoothed Particle Hydrodynamics (SPH) framework, producing dynamic density variations. Finally, a curved ray-tracing renderer visualizes these density changes as multi-view consistent optical distortions. 



\subsection{Thermally-Driven Dynamics with SPH}
\label{sec:thermalsph}
We model air as a collection of particles within an SPH framework, a mesh-free, Lagrangian method ideal for dynamic fluids. Each particle \(i\) carries physical properties including mass \(m_i\), position \(\mathbf{x}_i\), velocity \(\mathbf{v}_i\), density \(\rho_i\), and temperature \(T_i\): 
\begin{equation}
    i = \{m_i,~\mathbf{x}_i,~\mathbf{v}_i,~\rho_i,~T_i\}.
    \label{eq:particle}
\end{equation}
The fluid state at any point is interpolated from neighboring particles using a smoothing kernel \(W\), for which we employ Desbrun's spiky kernel~\cite{desbrun1996smoothed}.

Unlike prior work targeting high-fidelity offline simulation, our goal is to capture the essential dynamics of heat-induced turbulence at interactive rates. The process begins by updating each particle's temperature based on its interaction with heat sources and other particles. This new temperature field then generates forces that drive the fluid motion, as described below.
\subsubsection{Heat Transport Model}
\label{sec:heattransport}

Our heat transport system consists of two main types of interactions.

\paragraph*{1) Source-particle exchange:} Our simulation injects thermal energy into the fluid through one or more user-defined heat sources, which are modeled as hexahedral regions in space. The total heat $Q$ at position $\mathbf{x}$ with temperature $T$ during a single time step is the sum of the individual contributions from all sources present in the scene:
\begin{equation}
    Q(\mathbf{x},T)
    \;=\;
    \sum_{s} q_{s}(\mathbf{x},T),
    \label{eq:total_source_heat}
\end{equation}
where $q_{s}$ is the heat flux from an individual source~$s$.
The calculation of $q_{s}$ is designed to be localized and is fundamentally driven by the temperature difference between the source $T_s$ and the particle $T$. To ensure a smooth falloff, the model defines an "influence zone" of radius $R_{s}$ around each source, and the amount of heat transferred decays linearly with the distance from the source's boundary. The heat flux from source ~$s$ is therefore given by the piecewise function:
\begin{equation}
    q_{s}(\mathbf{x},T)
    =
    \begin{cases}
        (T_{s}-T)\,\lambda_{s}\,r\,\Delta t,
        & \begin{tabular}[t]{@{}l@{}}
            if $\mathbf{x}$ lies \emph{inside} \\
            the source,
          \end{tabular} \\[6pt]
        (T_{s}-T)\,\lambda_{s}\,r\,\Delta t\!\left(1-\dfrac{d_{s}}{R_{s}}\right),
        & \text{if } d_{s}<R_{s}, \\[6pt]
        0,
        & \text{otherwise.}
    \end{cases}
    \label{eq:source_particle_heat}
\end{equation}
where $d_{s}$ represents the shortest distance from the particle to the surface of source $s$, $\lambda_{s}$ is a user-defined influence coefficient controlling the source's strength, $r$ the heat-transfer rate between sources and particles, and $\Delta t$ the simulation time-step.

\paragraph*{2) Inter-particle conduction:} To model the diffusion of heat throughout the fluid,  we start from the continuous form of the Fourier heat-conduction equation \cite{CLEARY1999227}, which governs transient heat flow:
\begin{equation}
    \rho\,c\,\frac{\partial T}{\partial t}
    \;=\;
    \nabla\!\cdot\!\bigl(k\,\nabla T\bigr) \;+\; Q,
    \label{eq:fourier}
\end{equation}
where $\rho$ is the material density, $c$ the specific heat capacity, $T$ the temperature field, $t$ time, $k$ the thermal conductivity, and $Q$ is the energy from external heat sources. The operator $\nabla\!\cdot(\,\cdot\,)$ denotes the divergence, while $\nabla T$ is the temperature gradient.

To adapt this for our SPH particle system, we must discretize it. However, directly discretizing the second-order derivatives in $\nabla\!\cdot\!\bigl(k\,\nabla T\bigr)$ can lead to numerical instability. We therefore adopt the more robust pairwise heat-transfer formulation from~\cite{sphheat}, which models conduction as a one-dimensional heat exchange along the line connecting each particle pair $(i,j)$.  This approach enforces energy conservation and improves stability. The resulting SPH discretization for the rate of temperature change of particle $i$ due to conduction is: 
\begin{equation}
\frac{\partial T_i}{\partial t} =
\frac{1}{\rho_i c_i} \sum_j \frac{m_j}{\rho_j}
\cdot \frac{4 k_i k_j}{k_i + k_j}
\cdot \frac{\mathbf{r}_{ij} \cdot \nabla_i W_{ij}}{\|\mathbf{r}_{ij}\|^2}
\cdot (T_j - T_i),
\label{eq:sph_heat_transfer}
\end{equation}
where the sum is over all neighboring particles $j$. Here,  $m_j$ and $\rho_j$ are the mass and density of particle $j$, $k_i$ and $k_j$ are their thermal conductivities, $\mathbf{r}_{ij} = \mathbf{r}_i - \mathbf{r}_j$ is the vector between them, and $\nabla_i W_{ij}$ is the gradient of the smoothing kernel W.

For our specific case of simulating air, we introduce two simplifying assumptions to enhance interactive-rate performance. 
We assume a constant specific heat capacity $(c_i = 1)$ and a uniform thermal conductivity $k_i = k_j=k$ for all particles. This reduces the full equation to a more computationally efficient form:
\begin{equation}
\frac{\partial T_i}{\partial t} =
\frac{1}{\rho_i} \sum_j
\left[
\frac{2k m_j}{\rho_j}
\cdot \frac{\mathbf{r}_{ij} \cdot \nabla_i W_{ij}}{\|\mathbf{r}_{ij}\|^2}
\cdot (T_j - T_i)
\right].
\label{eq:simplified_heat_transfer}
\end{equation}

While not strictly physical, this is a deliberate design choice that plausibly enhances turbulent detail at a negligible computational cost, which is essential for achieving interactive-rate performance.

\subsubsection{Thermal Force Generation}
Once the temperature field is updated through conduction and source interactions, the next step is to translate these \textit{thermal changes} into \textit{particle motion}. Our system accomplishes this with a simplified thermal force model designed to be both computationally efficient and visually plausible. This model captures the two primary effects of heated air:

\textbf{1) Large-scale vertical motion:} The dominant effect is \textit{buoyancy force} $\mathbf{F}^{\mathrm{buoy}}$, which we model based on Archimedes’ principle:
\begin{equation}
    \mathbf{F}^{\mathrm{buoy}}
    \;=\;
    \bigl(\rho_{\mathrm{env}} - \rho_{\mathrm{s}}\bigr)\,
    g\,V,
    \label{eq:buoyancy_arch}
\end{equation}
where $(\rho_{\mathrm{env}}-\rho_{\mathrm{s}})$ is the density difference between the ambient and heated gases, $g$ is the gravitational acceleration, and $V$ is the volume. This equation is consistent with the Boussinesq approximation~\cite{boussinesq1902mise} and buoyancy forces have been used extensively for fluid simulation~\cite{turner1973buoyancy,fedkiw2001visual}. 

To make this efficient for our particle system, we use a linearized density–temperature relation, which allows us to express the buoyancy force directly in terms of temperature difference:
\begin{equation}
    \mathbf{F}^{\mathrm{buoy}}_i
    \;=\;
    K\,
    \bigl(T_{\mathrm{i}} - T_{\mathrm{env}}\bigr)\,
    g\,V.
    \label{eq:buoyancy_linear}
\end{equation}
This force is then used to compute the buoyancy-induced velocity change for each particle $i$ over the time step $\Delta t$:
\begin{equation}
\Delta \mathbf{v}^{\mathrm{buoy}}_i
= \frac{\mathbf{F}^{\mathrm{buoy}}_i}{m_i}\,\Delta t.
\label{eq:velocity_change}
\end{equation}

\textbf{2) Smaller-scale swirling motion:}
In addition to the large-scale motion from buoyancy, we introduce a \textit{convective velocity term} to enhance smaller-scale turbulent details. This is a deliberate design choice for visual plausibility, where we assume particles tend to accelerate toward regions of increasing temperature. This effect is modeled as a simple velocity increment:
\begin{equation}
\Delta \mathbf{v}^{\mathrm{temp}}_i = \beta\,\frac{\nabla T_i}{\|\nabla T_i\|},
\label{eq:thermal_convection}
\end{equation}
where $\beta$ is a user-defined multiplier that controls the strength of the effect. $\nabla T_i / \|\nabla T_i\|$ is the unit vector in the direction of steepest temperature increase. 
While not physically rigorous, this model provides an inexpensive way to add swirling motion that is crucial for the final visual appearance.

These two components are combined to produce a total thermal velocity update:
\begin{equation}
    \Delta \mathbf{v}^{\mathrm{thermal}}_i = \Delta \mathbf{v}^{\mathrm{buoy}}_i + \Delta \mathbf{v}^{\mathrm{temp}}_i.
    \label{eq:thermal}
\end{equation}
This total update is applied at the beginning of each simulation time step, serving as the primary driver for the fluid's motion before the SPH solver resolves the pressure and density constraints.

\begin{figure*}
    \centering
    \includegraphics[width=1\textwidth]{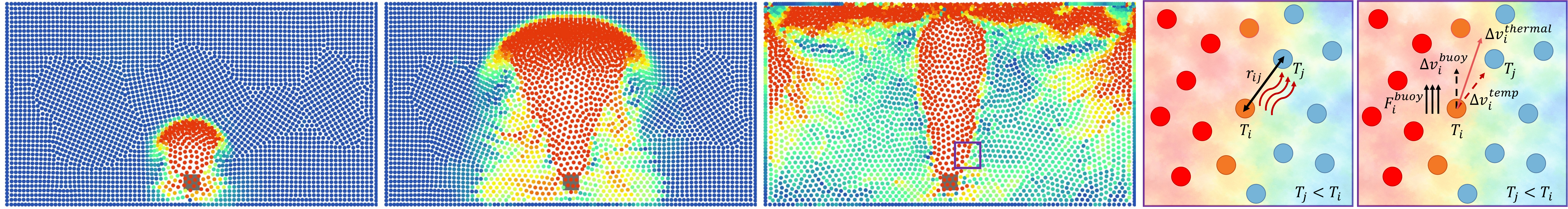}
    \caption{2D dissection of our heat-driven particle system. The bottom square marks the heat source. Once activated, particles absorb heat, rise, and spread turbulence the domain (left to right). Colors denote velocity magnitude, from slow (blue) to fast (red).}    
    \vspace{-0.5cm}
    \label{fig:moveheatbeamparticle}
\end{figure*}
\subsection{SPH Solver and State Update}
\label{sec:sphsolver}
Next, we apply the SPH solver to all particles, and update their state accordingly. Note that our thermal force model is designed to be modular and can be seamlessly integrated into \textit{any efficient SPH framework} as an external force update. For our implementation, we chose the recent semi-implicit SPH solver by He et al.~\cite{sissm} to serve as the computational backbone. 

At each simulation step, we perform three stages:
\textit{1) Thermal prediction (external update).}
We first update temperature via heat exchange and apply the resulting buoyancy/thermal acceleration to obtain predicted velocities and an intermediate position $\mathbf{x}_i^{*}$. This stage injects the heat-driven dynamics that generate plumes and turbulence.
\textit{2) SPH constraint solver (backbone).}
Starting from $\mathbf{x}_i^{*}$, we run the SPH solver to enforce the fluid constraints (e.g., constant density) by iteratively correcting particle positions, resolving pressure effects through the solver’s optimization procedure.
\textit{3) State update.}
Finally, we update the particle velocity from the net displacement over the time step, ensuring the state reflects both the thermal forcing (Stage~1) and the solver’s pressure corrections (Stage~2).


Fig.~\ref{fig:moveheatbeamparticle} illustrates this process: a localized heat source drives initially stationary particles into a rising convective plume (blue$\rightarrow$red velocity transition), dynamically reshaping the density field required for the subsequent refractive rendering.

\subsection{Volumetric Rendering of Optical Distortions}
\label{sec:adaptiveray}
Heat haze is visible because density variations in air induce a spatially varying refractive index, bending light rays and producing depth-dependent distortions. Our renderer therefore converts the simulated density field into a refractive-index volume and traces \emph{curved} rays through this inhomogeneous medium.

\subsubsection{Refractive Index from Density}
We first rasterize the particle densities into a regular 3D grid (resolution $\mathbf{R}$) and evaluate a continuous density field $\rho(\mathbf{x})$ via SPH kernel splatting at voxel centers. We then map density to refractive index using the Gladstone--Dale relation~\cite{gladstonedale}:
\begin{equation}
n(\mathbf{x}) = 1 + K\,\cdot \rho(\mathbf{x}),
\label{eq:gladstone_dale}
\end{equation}

where $K = \gamma\, \cdot K_{\mathrm{GD}}$, and $\gamma$ is a user-defined parameter that converts the SPH system density to the physical density scale. $K_{\mathrm{GD}}$ is the Gladstone--Dale constant ($K_{\mathrm{GD}} \approx 2.27 \times 10^{-4}\,\mathrm{m}^3/\mathrm{kg}$ for air at room temperature). In our system, we set $K = 2 \times 10^{-7}$ to produce turbulence effects that are visually consistent with real-world observations, while larger values are used to generate more pronounced refractive distortions for visualization. Crucially, the ray curvature is driven by the refractive-index gradient, which is directly proportional to the density gradient:
\begin{equation}
\nabla n(\mathbf{x}) = K\,\cdot \nabla \rho(\mathbf{x}).
\label{eq:gd_gradient}
\end{equation}
In practice, we compute $\nabla \rho$ on the grid using finite differences and sample both $n(\mathbf{x})$ and $\nabla n(\mathbf{x})$ along the ray with trilinear interpolation.


\begin{figure}[!htb]
    \centering
    \includegraphics[width=0.45\textwidth]{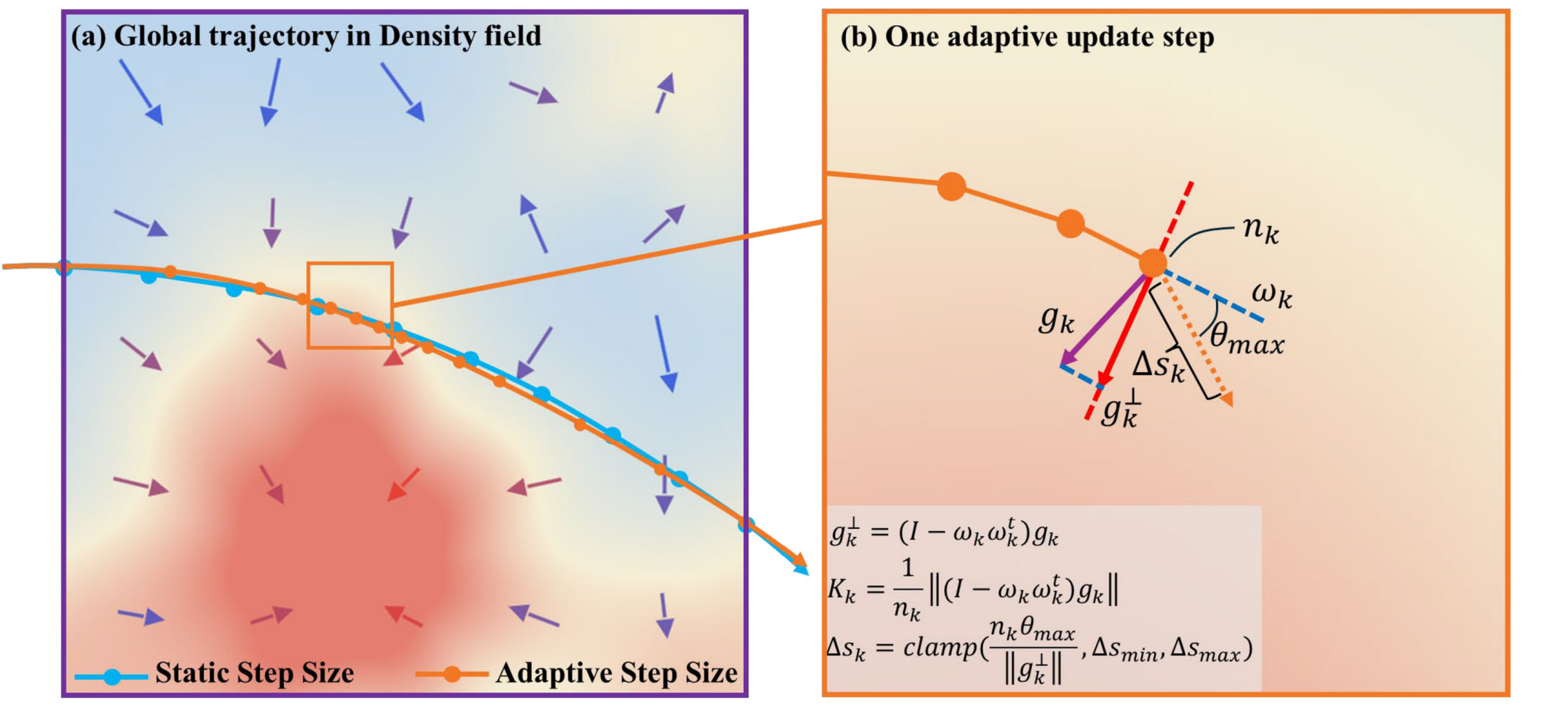}
    \caption{2D illustration of curved ray tracing with adaptive step size. Red regions indicate higher density, while blue regions indicate lower density. Our adaptive scheme places fewer samples where the density gradient is low and concentrates them where it is high, automatically balancing accuracy against cost.}
    \label{fig:2dcurvedraymarching}
\end{figure}

\subsubsection{Curved Ray Tracing Through the Refractive Volume}
Let a camera ray follow a curve $\mathbf{C}(s)\in\mathbb{R}^3$ parameterized by arc length $s$, with unit tangent (ray direction) $\boldsymbol{\omega}(s)=\frac{d\mathbf{C}}{ds}$ and $\|\boldsymbol{\omega}(s)\|=1$. In an inhomogeneous refractive medium, the ray path satisfies the standard vector ray equation (from Fermat/Eikonal optics):
\begin{equation}
\frac{d}{ds}\!\Big(n(\mathbf{C}(s))\,\boldsymbol{\omega}(s)\Big) = \nabla n(\mathbf{C}(s)).
\label{eq:ray_equation}
\end{equation}
Expanding Eq.~\eqref{eq:ray_equation} yields an ODE for the change in direction that depends only on the component of $\nabla n$ \emph{perpendicular} to the ray:
\begin{equation}
\frac{d\boldsymbol{\omega}}{ds}
=
\frac{1}{n(\mathbf{C}(s))}\Big(\mathbf{I}-\boldsymbol{\omega}\boldsymbol{\omega}^{\mathsf{T}}\Big)\,\nabla n(\mathbf{C}(s)).
\label{eq:ray_ode}
\end{equation}
We integrate Eq.~\eqref{eq:ray_ode} with a simple, stable explicit update. At step $k$ (position $\mathbf{C}_k$, direction $\boldsymbol{\omega}_k$), let $n_k=n(\mathbf{C}_k)$ and $\mathbf{g}_k=\nabla n(\mathbf{C}_k)$. We update
\begin{equation}
\begin{aligned}
\boldsymbol{\omega}_{k+1}
&=
\mathrm{normalize}\!\left(
\boldsymbol{\omega}_k
+
\frac{\Delta s_k}{n_k}\Big(\mathbf{I}-\boldsymbol{\omega}_k\boldsymbol{\omega}_k^{\mathsf{T}}\Big)\mathbf{g}_k
\right), \\
\mathbf{C}_{k+1}
&=
\mathbf{C}_k+\Delta s_k\,\boldsymbol{\omega}_{k+1}.
\end{aligned}
\label{eq:ray_discrete}
\end{equation}
Because clear air is effectively non-participating, we ignore absorption/scattering/emission and generate the image purely from ray bending: each pixel samples the scene radiance along the refracted ray via scene intersection.

\subsubsection{Adaptive Step Size for Efficient and Stable Ray Integration}
A fixed step size wastes work in smooth regions and can miss high-frequency distortions near sharp refractive-index gradients (e.g., near plume boundaries). We therefore choose $\Delta s_k$ adaptively using the local \emph{angular change rate} of the ray direction. From Eq.~\eqref{eq:ray_ode}, the local curvature magnitude is
\begin{equation}
\kappa_k
=
\left\lVert \frac{d\boldsymbol{\omega}}{ds} \right\rVert
=
\frac{1}{n_k}\left\lVert
\Big(\mathbf{I}-\boldsymbol{\omega}_k\boldsymbol{\omega}_k^{\mathsf{T}}\Big)\mathbf{g}_k
\right\rVert.
\label{eq:curvature}
\end{equation}
We then bound the maximum turning angle per step by a user-specified tolerance $\theta_{\max}$ (in radians) and compute
\begin{equation}
\label{eq:adaptive_stepsize}
\begin{split}
\Delta s_k
&=
\mathrm{clamp}\!\left(
\frac{\theta_{\max}}{\kappa_k+\varepsilon},
\;\Delta s_{\min},\;\Delta s_{\max}
\right) \\
&=
\mathrm{clamp}\!\left(
\frac{n_k\,\theta_{\max}}{
\left\lVert
\Big(\mathbf{I}-\boldsymbol{\omega}_k\boldsymbol{\omega}_k^{\mathsf{T}}\Big)\mathbf{g}_k
\right\rVert+\varepsilon},
\;\Delta s_{\min},\;\Delta s_{\max}
\right),
\end{split}
\end{equation}
where $\varepsilon$ avoids division by zero and $\Delta s_{\min},\Delta s_{\max}$ bound the step length for robustness. 
Intuitively, by Eq.~\eqref{eq:gd_gradient} the refractive-index gradient scales with the density gradient, so regions with large $\|\nabla\rho\|$ induce higher ray curvature (Eq.~\eqref{eq:curvature}) and therefore require smaller steps in Eq.~\eqref{eq:adaptive_stepsize} to bound the per-step angular error, as shown in Fig. ~\ref{fig:2dcurvedraymarching}. Pseudocode is provided in the
supplemental material.

\section{Experimental Results}
\label{sec:results}

We evaluate our system along three axes: (1) \textbf{interactive performance} and scalability, (2) \textbf{physical plausibility} of volumetric heat-haze distortions, and (3) our key capability: \textbf{multi-view consistent} rendering from a single 3D turbulence volume. We compare to representative image-based turbulence simulators and validate against a controlled real capture. 


\subsection{Implementation and Settings}


 We implement SPH simulation in Unity compute shaders and curved ray marching in fragment shaders. All experiments run on an NVIDIA RTX~4090 (24\,GB) with an Intel i9-13900K. Unless otherwise specified, we use $\Delta t{=}0.006$s and $N\in[0.5,1.5]$ M particles for interactive frame rates. 
Key parameters include thermal conductivity $k=100$, buoyancy constant $K=100$, convection multiplier $\beta=50$, smoothing radius $H=0.1$, 
and SPH reference density $\rho_0=300$.
For completeness and stress-testing, we also support a fixed-step variant by setting $\Delta s_k\equiv \Delta s$. A full list is provided in a table in the supplemental material. The simulation domain is closed and models a local atmospheric region around a heat source. Particles rebound at the domain boundary and are reset to ambient temperature there, approximating thermal dissipation and preventing unbounded heating. For efficiency, particles do not collide with scene geometry and may enter opaque objects; since rendering rays terminate at the first opaque surface, interior density never affects the image.


\textbf{Scalability:}
Tab.~\ref{tab:scalability_adaptive_vs_static_compact} reports scalability with respect to particle count, simulation volume size, and number of simultaneous camera views. Each entry shows \textbf{Adaptive/Static} step-size curved-ray tracing (static $\Delta s {=} 0.02$), demonstrating that adaptive stepping substantially improves throughput while retaining the same volumetric model.



\begin{table}[t]
  \centering
  \small
  \caption{Mean FPS on an NVIDIA RTX~4090 (24\,GB) at 1920$\times$1080. Each entry reports \textbf{Adaptive/Static} step-size curved-ray tracing (static $\Delta s{=}0.02$). Baseline (when not varied): $N{=}1.5$M, simulation volume $10^3$ (m³), 1 camera.}
  \label{tab:scalability_adaptive_vs_static_compact}
  \setlength{\tabcolsep}{4pt}
  \begin{tabular}{@{}l|ccccc@{}}
    \toprule
    \multicolumn{6}{c}{\textbf{Particles $N$ (millions)}} \\
    \hline
    Values & 0.50 & 1.0 & 1.5 & 2.00 & 2.5 \\
    \hline
    FPS $\uparrow$ & 75.5/50.6 & 49.3/35.1 & 35.1/25.5 & 25.7/20.5 & 18.9/17.5 \\
    \toprule
    \multicolumn{6}{c}{\textbf{Simulation Volume (m³)}} \\
    \hline
    Values & $10^3$ & $15^3$ & $20^3$ & $25^3$ & $30^3$ \\
    \hline
    FPS $\uparrow$ & 35.1/25.5 & 41.2/25.0 & 43.0/25.4 & 42.2/23.9 & 43.5/22.6 \\
    \toprule
    \multicolumn{6}{c}{\textbf{\# Cameras}} \\
    \hline
    Values & 1 & 2 & 3 & 4 & 5 \\
    \hline
    FPS $\uparrow$ & 35.1/25.5 & 27.0/20.3 & 23.4/16.4 & 21.2/13.8 & 20.2/12.5 \\
    \bottomrule
  \end{tabular}
  \vspace{-0.5cm}
\end{table}

\begin{figure}[!htb]
    \centering
    \includegraphics[width=0.45\textwidth]{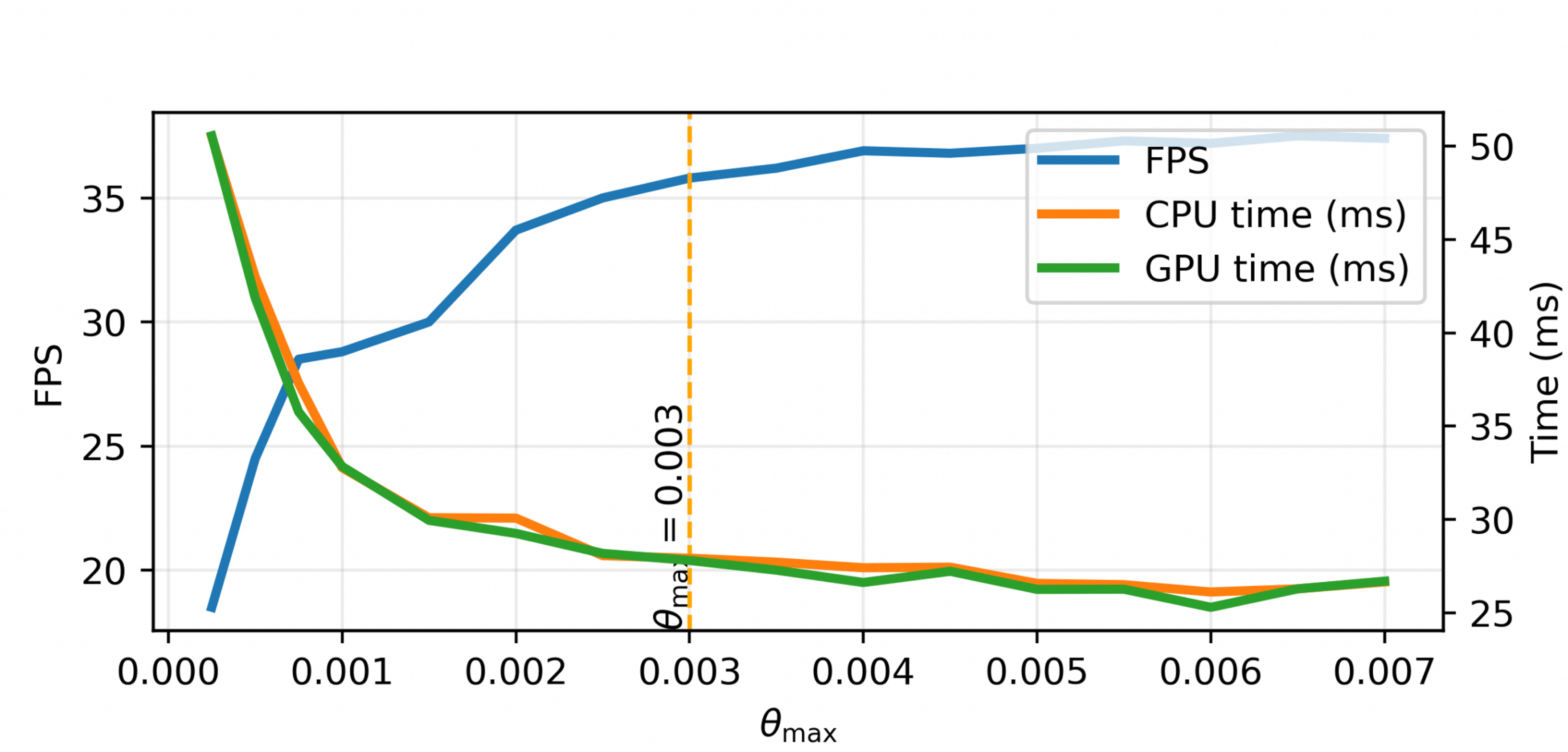}
    \caption{FPS, CPU time, and GPU time versus the error tolerance $\theta_{\max}$.
    }
    \label{fig:analyzetheta}
\end{figure}

\begin{figure}[!htb]
    \centering
    \includegraphics[width=0.45\textwidth]{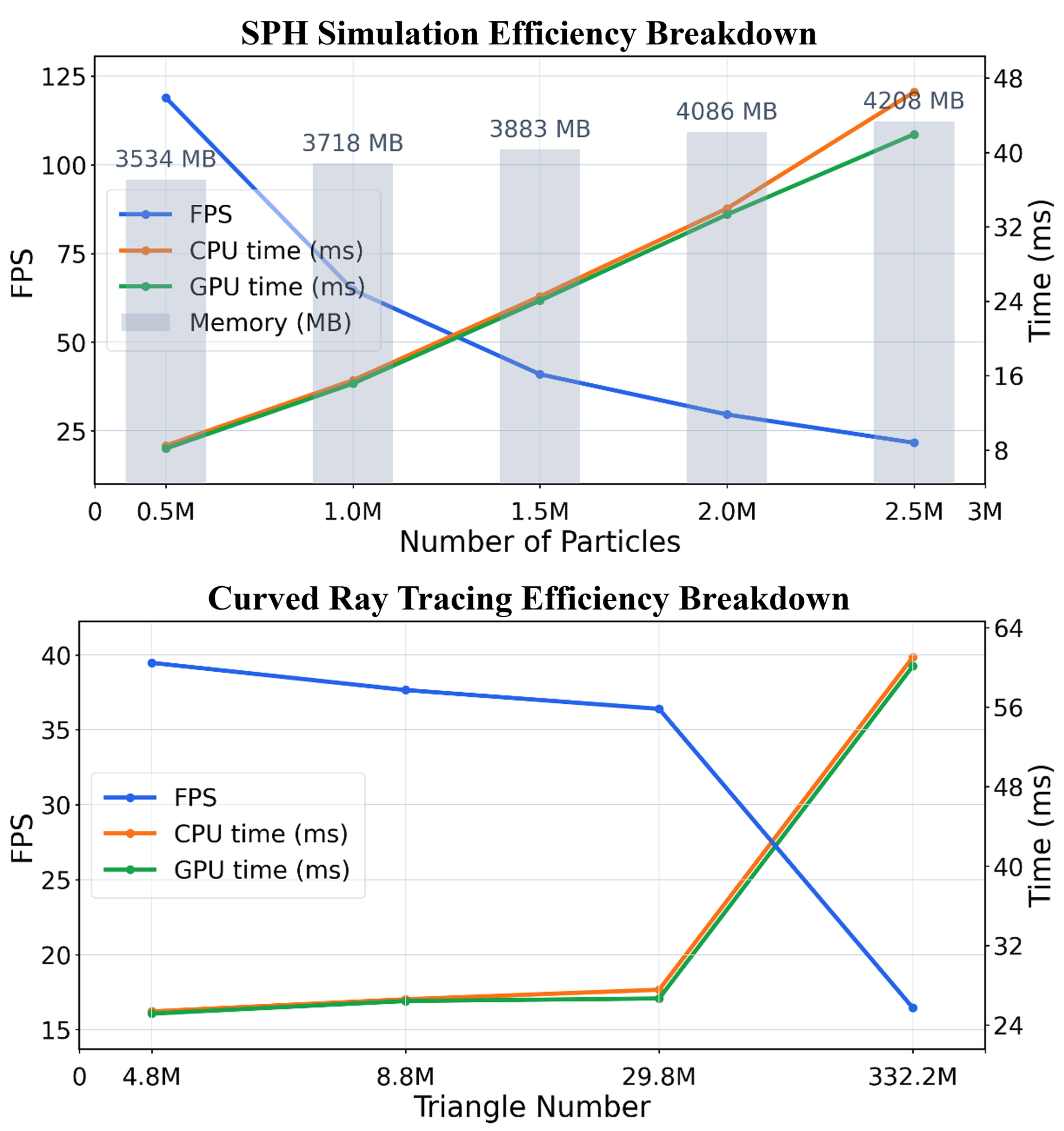}
    \caption{Performance breakdown for the SPH simulation and curved-ray-tracing components evaluated separately.
    }
    \vspace{-0.3cm}
    \label{fig:analyzebreakdown}
\end{figure}


\subsection{Latency}
\begin{figure*}[!]
    \centering
    \includegraphics[width=1\textwidth]{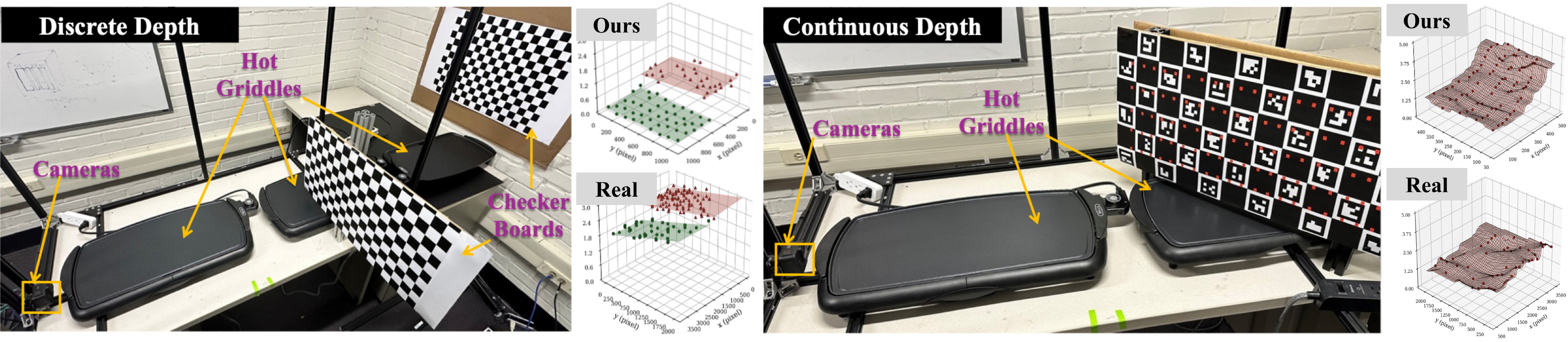}
    \vspace{-10pt}
    \caption{Under constant turbulence strength, the angle-of-arrival variance grows linearly with distance. We validate this trend in two configurations: (1) {Discrete depths} with two parallel checkerboards at different $L$ (farther plane yields larger variance), and (2) {Continuous depth} using a slanted checkerboard, where variance increases approximately linearly along the depth gradient. Our simulated results match the trend observed in the real captures (More results in Fig.~\ref{fig:allexperiment}).}    
    \label{fig:real-setting}
    \vspace{-0.5cm}
\end{figure*}




\textbf{Adaptive tolerance sweep:}
Fig.~\ref{fig:analyzetheta} shows the accuracy--speed trade-off controlled by the maximum turning angle per step $\theta_{\max}$ (Eqn.~\ref{eq:adaptive_stepsize}). Increasing $\theta_{\max}$ increases the mean step size and improves FPS until diminishing returns; we set $\theta_{\max}=0.003$ for all adaptive experiments.

\textbf{Breakdown analysis of simulation and rendering:}
We analyze the performance of SPH simulation and curved ray tracing separately by measuring FPS, CPU time, and GPU time, as shown in Fig.~\ref{fig:analyzebreakdown}. The entire pipeline runs on the GPU as Unity compute and fragment shaders. Times are measured with the Unity Profiler, and CPU time mainly reflects engine-side orchestration, not simulation. With 1.5M particles, a \(10^3\,\mathrm{m}^3\) simulation volume, one camera, and a checkerboard target, heat transport accounts for 5.9\% (1.5 ms) of GPU time, SPH updates for 74.0\% (18.8 ms), and rendering for 19.9\% (5.0 ms). Increasing the number of SPH particles improves simulation fidelity but leads to higher computational cost and lower FPS. For rendering, we evaluate performance across scenes with varying triangle counts using adaptive curved ray tracing. As scene complexity increases, the cost rises and performance drops, while interactive frame rates are still maintained for scenes with up to 332.2M triangles.



\subsection{Qualitative Results}

\begin{figure*}
    \centering
    \includegraphics[width=01\textwidth]{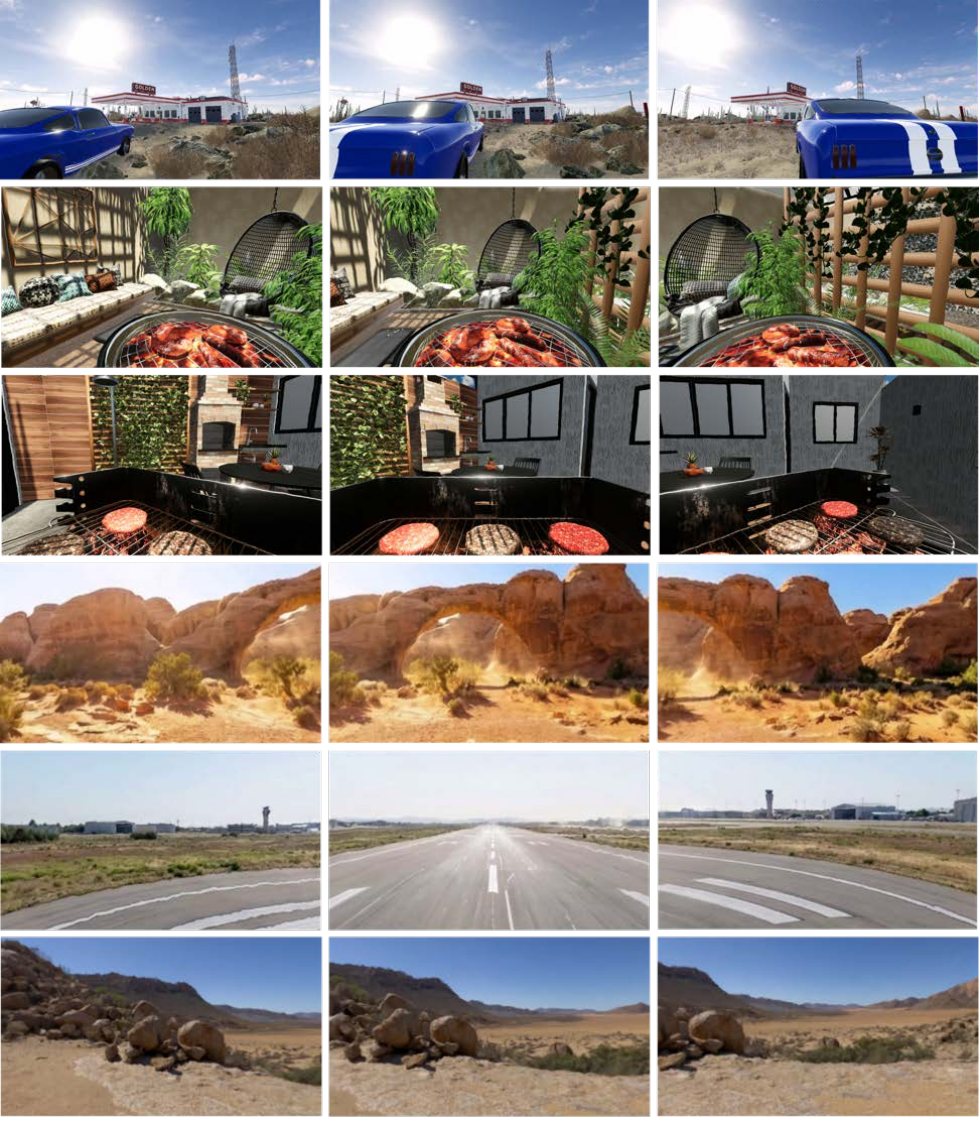}
    \caption{Simulation and rendering results of our heat-induced turbulence. Turbulence strength is set higher than typical in order to highlight the distortion effects. Top three rows: renderings from a fixed camera over time. Bottom three rows: the same simulation rendered from varying camera positions and orientations. More video results can be seen in the supplemental material.}    
    \label{fig:results}
\end{figure*}

We rendered six different scenes to qualitatively demonstrate the atmospheric turbulence effects generated by our method, as shown in Fig.~\ref{fig:results}. We first chose three scenes with complex geometry, textures, and lighting conditions, containing 332.2M, 8.61M and 29.83M triangles, respectively. We placed a desert gas station scene and two yard scenes featuring a grill stove, with the camera positioned to observe each scene through turbulence. In addition, we selected three real-world HDRI environments to produce turbulence effects that can be directly compared with real captured videos. In all scenes, the camera is movable, allowing turbulence to be observed from varying viewpoints. We also provide one scene with a dynamic object and a moving camera as shown in Fig. S11 in supplemental material. In contrast to image-based turbulence simulators, which apply turbulence independently to each frame and often produce discontinuous artifacts, our volumetric simulation yields a spatially and temporally continuous turbulence field, enabling smooth rendering across viewpoints and over time. More results are included in the supplementary video.

\subsection{Validation and Comparative Analysis on Real Experimental Data}

\begin{figure*}[!t]
    \centering
    \includegraphics[width=1\textwidth]{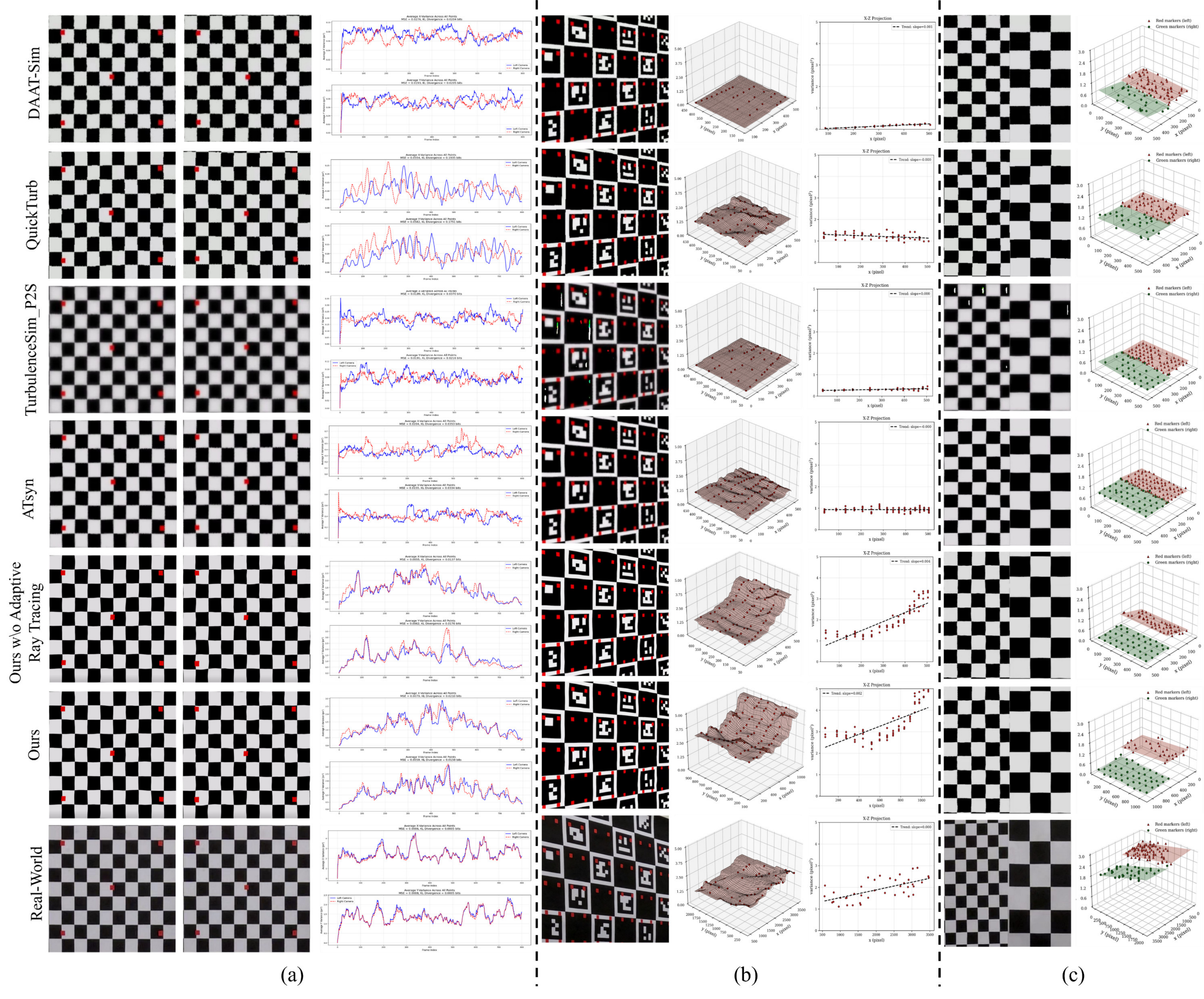}
    \caption{Analysis of our method compared with baselines and real-world results (zoom in for details, enlarged plots are in supplemental material). 
(a) \textbf{Multi-View Consistency:} two cameras with a small baseline capture turbulence-affected imagery; variance curves of tracked points along $x$/$y$ axes from our simulator overlap closely between cameras, matching real data. 
(b) \textbf{Continuous Depth:} with a tilted plane, variance projected onto the $x$--$z$ plane increases with distance, consistent with real-world trends. 
(c) \textbf{Discrete Depths:} with two parallel planes at different depths, simulator results show variance clustered on two separated planes, in agreement with real observations.
}
    \label{fig:allexperiment}
    \vspace{-0.5cm}
\end{figure*}

\begin{figure*}
    \centering
    \includegraphics[width=1\textwidth]{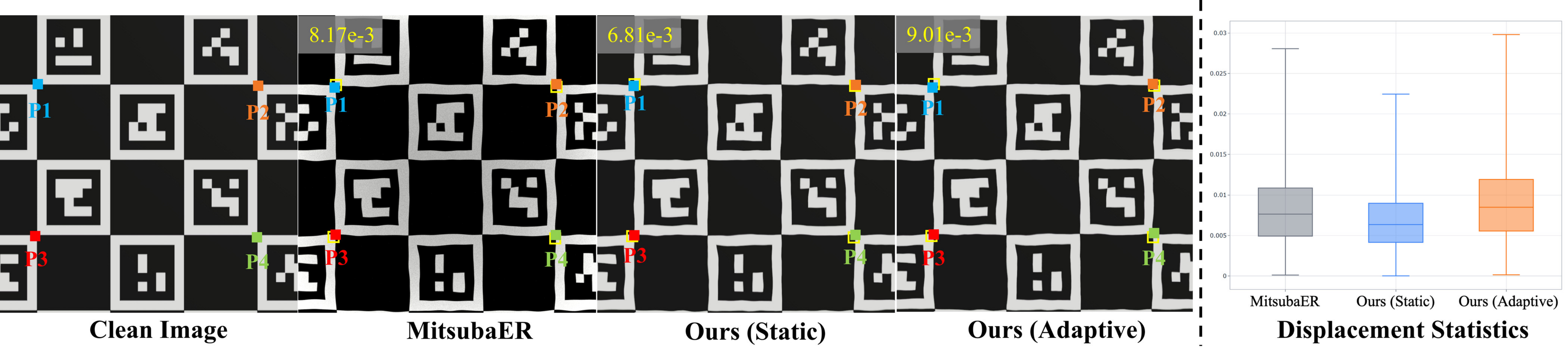}
    \caption{Qualitative comparison between our simulator and MitsubaER. Colored squares show ray–image plane intersections from uniformly sampled 3D points, while yellow squares indicate reference intersections from straight ray tracing.}
    \label{fig:comparewithoffline}
    \vspace{-0.5cm}
\end{figure*}

\textbf{Overview:} We compare against representative image-based turbulence simulators: DAAT-Sim~\cite{daatsim}, QuickTurb~\cite{saha2024turb}, TurbulenceSimP2S~\cite{mao2021accelerating}, and ATsyn~\cite{zhang2024spatio}. Note that these methods operate in 2D (screen-space), and cannot enforce \emph{geometrically consistent} distortions across viewpoints. In contrast, our method renders from a persistent 3D refractive-index field.
Since no prior work provides an interactive \emph{3D multi-view} heat-haze simulator with a shared refractive volume, these baselines represent the closest alternatives commonly used in practice.
However, we further evaluate the accuracy of our rendering results by additionally comparing our method against the offline renderer MitsubaER~\cite{pediredla2020path}, an open-source curved ray tracer based on bidirectional path tracing (BDPT). It updates curved ray trajectories using a user-specified fixed step size and a leapfrog-style integration scheme. This comparison allows us to validate that our adaptive curved ray tracing (performed at interactive rates) can be as accurate as an offline Monte Carlo-based renderer.

\textbf{Real-capture benchmark:}
We adopt a controlled laboratory setup similar to that of Tian et al.~\cite{tian2012depth}, as shown in Fig.~\ref{fig:real-setting}. Three adjacent electric cooking griddles, each 52cm $\times$ 26cm, are heated to 400\textdegree{F} to generate a consistent and controllable volume of thermal turbulence. A reference checkerboard scene is observed through this volume, a setup which allows for the emulation of various outdoor turbulence strengths equivalent to a path-length of several kilometers. To analyze the distortion, video sequences of the scene are captured at 30 fps. We captured videos using two GoPro HERO11 Black Mini cameras at 5.3K resolution with an aperture of f/2.5, operating in Linear mode to avoid lens edge distortion. 

Tian et al.~\cite{tian2012depth} reported a depth-dependent phenomenon in atmospheric turbulence, showing that the variance on the image plane is approximately proportional to the depth of the object from the camera. We adopt this finding as a benchmark to evaluate whether our simulator reproduces the depth effect correctly. We further generalize this principle to stereo imaging, which, to the best of our knowledge, has not been investigated in prior simulators: when multiple cameras simultaneously observe a scene through the same turbulence field from nearby yet distinct positions, the resulting distortions should exhibit geometric consistency across views. 

\textbf{Depth-dependent distortion (volumetric property):} A key advantage of our 3D simulation is its ability to correctly model how optical distortions accumulate over distance. This volumetric property, as discussed in~\cite{tian2012depth}, can be viewed as a depth-dependent distortion effect, which is absent in screen-space methods. Following the physical model in~\cite{tian2012depth}, the variance $\langle \alpha^2 \rangle$ of the angle-of-arrival fluctuations from a scene point at distance $L$ from the camera, under a constant turbulence strength $C_{n}^{2}$, is:
\begin{equation}
\label{eq:variance}
\begin{aligned}
\langle \alpha^2 \rangle
&= 2.914\, D^{-1/3} C_{n}^{2} \int_{0}^{L} \left(\frac{z}{L}\right)^{5/3}\, dz ~= \frac{3}{8} K_{n}^{2} L,
\end{aligned}
\end{equation}
where $K_{n}^{2} = 2.914\, D^{-1/3} C_{n}^{2}$ and $D$ denotes the aperture diameter.
We validate that our simulator adheres to this physical principle in two settings:

\textit{1) Discrete Depths:} We place two checkerboards at different distances from the camera (Fig.~\ref{fig:real-setting}). By tracking markers on each board, we measure the variance of their apparent motion. The results show two distinct clusters of variance, with the farther checkerboard exhibiting significantly higher variance, consistent with the theoretical model and real-world observations.

\textit{2) Continuous Depth:} We place a single checkerboard slanted with respect to the camera plane. As shown in Fig.~\ref{fig:real-setting}, the measured variance of markers on the board increases linearly with their distance from the camera, confirming that our simulation correctly integrates distortion along the view path. 
We conducted the same experiments with DAAT-Sim, QuickTurb,
TurbulenceSimP2S, and ATsyn, as shown in
Fig.~\ref{fig:allexperiment}. Among these methods, only our approach
reproduces the expected depth-dependent variance behavior: the variance
increases with continuously varying depth and forms clearly separated
clusters in the discrete-depth setting, consistent with the real-world
ground truth. The linear variance-depth prediction assumes constant
turbulence strength along each ray. Because our localized plume violates
this assumption, path-integrated turbulence strength and propagation
distance jointly produce the slightly U-shaped profile.

\textbf{Multi-view consistency:} The central contribution of our work is the ability to render a single, coherent turbulence volume from multiple viewpoints simultaneously. Unlike 2D image-based simulators that apply an independent, random distortion to each frame, our simulator's use of a persistent 3D density field and curved ray tracing naturally preserves multi-view coherence.
Specifically, we place two cameras with a small baseline observing the same turbulence volume and track checkerboard markers in both views, in the first three columns of Fig.~\ref{fig:allexperiment}. We plot variance curves of tracked points along x/y axes and we compare the cross-view displacement distributions using KL divergence as well as mean-squared error for these curves. The quantitative results are reported in Supplemental Tab. S2. Our method achieves substantially lower errors than the image-based baselines and approaches the consistency of the real capture.

\subsection{Validation for Curved Ray Tracing}
A dense real-world IoR field cannot be recovered from video, nor can an identical physical field be reproduced experimentally, so no real capture can serve as component-level ground truth for curved-ray integration. We therefore freeze one simulated IoR field and render it with both our interactive tracer and MitsubaER~\cite{pediredla2020path}, 	an offline curved-ray renderer designed for gradient-index materials such as lenses, which does not natively support atmospheric turbulence, under identical settings. 
For a fair comparison, we first captured a turbulence field from our simulator and reproduced the same scene setup in MitsubaER. The simulation domain is a cubic space with an edge length of 5 meters. We then rendered the field using both MitsubaER and our method with a checkerboard background, setting the step size of both methods to 0.01. We evaluated the results from two perspectives. 

\paragraph*{1) Per-ray trajectory comparison:} we uniformly sampled 1,000 rays from a frame and recorded their 3D positions at volume entry, volume exit, and every integration step inside the volume. For each ray pair, we report a mean per-step positional error of \(1.13 \times 10^{-3}\,\mathrm{m}\), a mean maximum positional error of \(1.27 \times 10^{-2}\,\mathrm{m}\), and a mean exit-position error of \(6.01 \times 10^{-3}\,\mathrm{m}\). The positional and exit-position errors are on the order of $10^{-3}$, while the maximum error is on the order of $10^{-2}$, suggesting that our ray trajectories closely match those of the offline renderer. 

\paragraph*{2) Visual pattern comparison:} We compared the rendered image patterns and performed a displacement analysis, as shown in Fig.~\ref{fig:comparewithoffline}. For this analysis, we uniformly sampled 18,225 3D points on the checkerboard plane and traced one ray per point toward the camera center. All rays were propagated through the same turbulence field, and we recorded the 3D positions of their intersections with the image plane. Based on these intersection points, we computed displacement between renders and straight ray tracing and visualized the statistics using box plots.
The average displacement from the straight-ray reference is
\(8.17\times10^{-3}\,\mathrm{m}\) for MitsubaER,
\(6.81\times10^{-3}\,\mathrm{m}\) for our static method, and
\(9.01\times10^{-3}\,\mathrm{m}\) for our adaptive method.
\begin{figure*}[t]
\includegraphics[width=\textwidth]{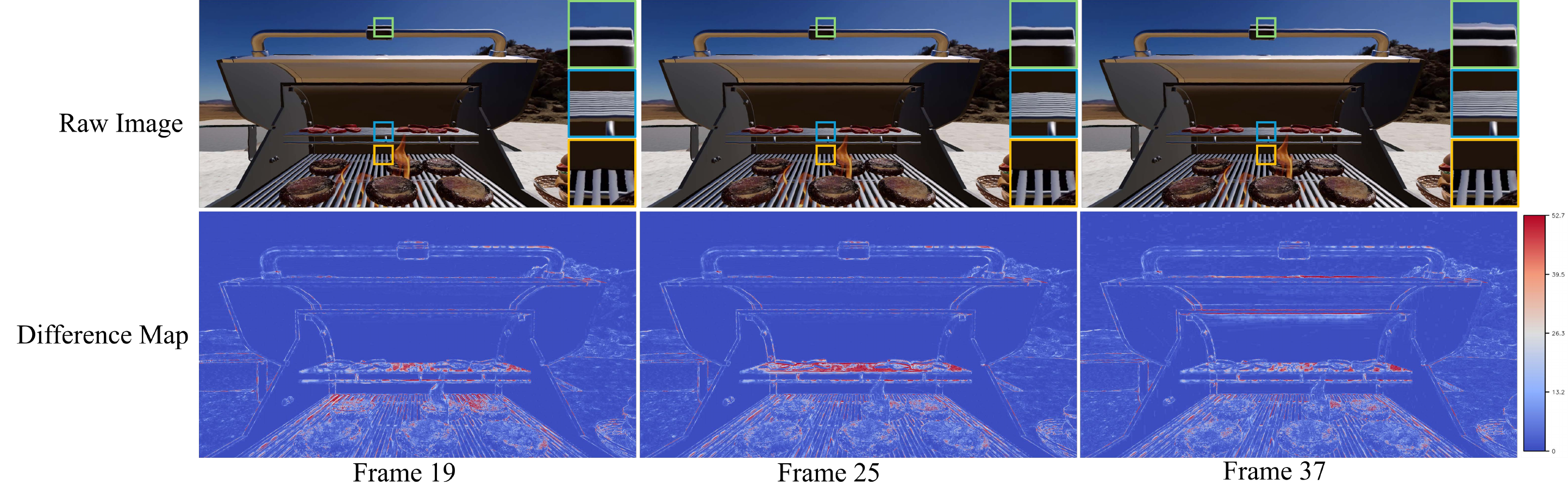}
\caption{The first row shows the raw RGB-rendered images, while the second row shows the corresponding difference maps with respect to the first frame of the video. A turbulence wave can be observed propagating from bottom to top as the stove is turned on. Please refer to the supplemental video for a clearer visualization of this effect.}
\label{fig:vs_perlin_bbq}
\vspace{-0.5cm}
\end{figure*}

\subsection{Comparison to 3D Perlin Noise}
We distinguish our method from 3D Perlin noise \cite{perlin2002improving}. Although 3D Perlin noise combined with curved ray tracing can produce volumetric, multi-view-consistent distortions with depth variation, it is a prescribed spatial field independent of heat and fluid dynamics. Modeling heat–turbulence interactions therefore requires additional condition-specific design, whereas our method couples these processes directly.
We evaluate this distinction from two perspectives.

\paragraph*{1) Turbulence distribution with heat:} We first initialize the system with the heat source turned off, such that the particles diffuse only due to the SPH system's tendency to reach the target density. After the system reaches a steady state, we activate a heat source at the bottom. The heat input causes rapid expansion of the nearby air and subsequently induces a bottom-to-top turbulence wave as the heated air rises, as shown in Fig.~\ref{fig:vs_perlin_bbq} and Fig. S8 in the supplemental material. We further visualize the temperature and IoR-gradient distributions using a small, thin square plane as the heat source by slicing and color-coding the corresponding volumes, as shown in Fig. S9 in the supplemental. We clamped the IoR values to 1.0003 to avoid physically unrealistic values. The temperature distribution naturally forms a torch-like rising plume, while larger IoR gradients appear in regions with stronger temperature differences. 

\paragraph*{2) Kolmogorov Scaling Analysis:} We further evaluate whether the spatial statistics of the generated IoR field are consistent with the Kolmogorov scaling law. Specifically, the refractive-index structure function is defined as $D_n(r)=\left\langle [n(\mathbf{x}+\mathbf{r})-n(\mathbf{x})]^2 \right\rangle$, which follows $D_n(r)=C_n^2 r^{2/3}$ under the Kolmogorov model. We therefore fit the measured structure function in log--log space and evaluate the slope error as $E_{\alpha}=\left|\alpha-\frac{2}{3}\right|$, where $\alpha$ is the fitted scaling exponent. For each of 300 simulated IoR fields, we construct a corresponding 3D Perlin-noise field with matched minimum, maximum, and mean IoR values to focus the comparison on spatial structure rather than the overall IoR range. Over the spatial range $r=0.5$--$2.5\,\mathrm{m}$, our method achieves $E_{\alpha}=0.3358$, compared with $0.6670$ for 3D Perlin noise, indicating that our simulated IoR field exhibits spatial scaling closer to the expected Kolmogorov behavior within this range.

\subsection{Ablation Study}
We evaluate the thermal forcing terms using mean vorticity
magnitude, which captures rotational (swirling) content associated with turbulent eddies~\cite{pope2001turbulent}, with its SPH definition provided in the supplemental
material. Buoyancy-only and convection-only produce mean
vorticities of 8.078 and 8.681, respectively, while the full model
achieves 8.919. These results indicate that buoyancy provides
large-scale vertical motion, whereas the convection term enriches
smaller-scale rotational structure.




To further compare the rendering quality of the static and adaptive step-size strategies, we decouple rendering error from simulation error by rendering the same checkerboard scene through the same turbulence field with different marching schemes. We compare the output of our adaptive marcher with that of a fixed small-step reference under the same refractive field on a tilted depth-varying checkerboard view. With the fixed step size set to 0.02, the two fields of view both achieve PSNR values around 45 and SSIM values of 0.99, with mean pixel error below 1: specifically, 45.64 PSNR, 0.999 SSIM, and 0.50 px mean pixel error under a 48-degree field of view, and 44.27 PSNR, 0.999 SSIM, and 0.39 px mean pixel error under a 55-degree field of view. This indicates that the adaptive step-size strategy yields results highly consistent with dense static sampling, while offering a better efficiency--accuracy trade-off.

\section{Conclusion}
We present the first interactive-rate simulator that couples thermally-driven SPH with curved-ray tracing to produce physically grounded, multi-view-consistent optical turbulence. Our framework models heat exchange between sources and particles, inter-particle thermal transport, and the resulting density-dependent refraction within a shared 3D volume. 

Our current system adopts several simulation-scale approximations, including temperature-driven velocity updates, the Gladstone--Dale relation, a linearized ideal-gas model, and simplified heat-source interactions. Higher-fidelity formulations could extend the method toward scientific-grade simulation and long-range ($>1\,\mathrm{km}$) imaging with long-focal-length cameras. The shared refractive volume also provides a useful platform for evaluating multi-view stereo under optical turbulence. Because our renderer terminates rays at the first opaque surface, it does not currently support transparent objects. Extending the framework to transparent scenes and other thermally driven fluids, such as water, offers promising directions for future work.
\section{Acknowledgments}
This work was supported by the following grants: W.Y. and N.L. were supported in part by NSF Award IIS-2232298 and OIA-2242812. S.J. was supported in part by NSF Award IIS-2232299. E.C., M.L., and H.C. were supported in part by NSF Award OIA-2531897, the NVIDIA Academic Grant Program, and a Google Research Gift. J. Y. was supported in part by NSF IIS-2232300.

\bibliographystyle{eg-alpha-doi} 
\bibliography{sample-base}       

@String{Computing = "Computing" }

@String{Computer = "{IEEE} Computer" }

@String{Springer = "Springer-Verlag" }

@article{yu2024adaptive,
  title={Adaptive ray tracing in freeform gradient-index media using an index directional derivative},
  author={Yu, Caiyun and Zong, Yi and Duan, Mingliang and Chen, Lei and Li, Jianxin},
  journal={Optics Letters},
  volume={49},
  number={19},
  pages={5375--5378},
  year={2024},
  publisher={Optica Publishing Group}
}

@article{pope2001turbulent,
  title={Turbulent flows},
  author={Pope, Stephen B},
  journal={Measurement Science and Technology},
  volume={12},
  number={11},
  pages={2020--2021},
  year={2001}
}

@inproceedings{liu2021turbulent,
  title={Turbulent details simulation for SPH fluids via vorticity refinement},
  author={Liu, Sinuo and Wang, Xiaokun and Ban, Xiaojuan and Xu, Yanrui and Zhou, Jing and Kosinka, Ji{\v{r}}{\'\i} and Telea, Alexandru C},
  booktitle={Computer Graphics Forum},
  volume={40},
  number={1},
  pages={54--67},
  year={2021},
  organization={Wiley Online Library}
}

@article{zhao2006visual,
  title={Visual simulation of heat shimmering and mirage},
  author={Zhao, Ye and Han, Yiping and Fan, Zhe and Qiu, Feng and Kuo, Yu-Chuan and Kaufman, Arie E and Mueller, Klaus},
  journal={IEEE Transactions on Visualization and Computer Graphics},
  volume={13},
  number={1},
  pages={179--189},
  year={2006},
  publisher={IEEE}
}

@article{jeong2003smoothed,
  title={Smoothed particle hydrodynamics: Applications to heat conduction},
  author={Jeong, JH and Jhon, MS and Halow, JS and Van Osdol, J},
  journal={Computer Physics Communications},
  volume={153},
  number={1},
  pages={71--84},
  year={2003},
  publisher={Elsevier}
}

@article{bender2018turbulent,
  title={Turbulent micropolar SPH fluids with foam},
  author={Bender, Jan and Koschier, Dan and Kugelstadt, Tassilo and Weiler, Marcel},
  journal={IEEE Transactions on Visualization and Computer Graphics},
  volume={25},
  number={6},
  pages={2284--2295},
  year={2018},
  publisher={IEEE}
}

@article{pediredla2020path,
  title={Path tracing estimators for refractive radiative transfer},
  author={Pediredla, Adithya and Chalmiani, Yasin Karimi and Scopelliti, Matteo Giuseppe and Chamanzar, Maysamreza and Narasimhan, Srinivasa and Gkioulekas, Ioannis},
  journal={ACM Transactions on Graphics (TOG)},
  volume={39},
  number={6},
  pages={1--15},
  year={2020},
  publisher={ACM New York, NY, USA}
}

@ArtifactSoftware{R,
    title = {R: A Language and Environment for Statistical Computing},
    author = {{R Core Team}},
    organization = {R Foundation for Statistical Computing},
    address = {Vienna, Austria},
    year = {2019},
    url = {https://www.R-project.org/},
}

@article{lagae2010survey,
  title={A survey of procedural noise functions},
  author={Lagae, Ares and Lefebvre, Sylvain and Cook, Rob and DeRose, Tony and Drettakis, George and Ebert, David S and Lewis, John P and Perlin, Ken and Zwicker, Matthias},
  journal={Computer Graphics Forum},
  volume={29},
  number={8},
  pages={2579--2600},
  year={2010},
  organization={Wiley Online Library}
}

@article{perlin85,
author = {Perlin, Ken},
title = {An Image Synthesizer},
year = {1985},
issue_date = {Jul. 1985},
publisher = {Association for Computing Machinery},
address = {New York, NY, USA},
volume = {19},
number = {3},
issn = {0097-8930},
url = {https://doi.org/10.1145/325165.325247},
doi = {10.1145/325165.325247},
journal = {SIGGRAPH Comput. Graph.},
month = {jul},
pages = {287–296},
numpages = {10}
}

@ARTICLE{chimitt2022,
  author={Chimitt, Nicholas and Zhang, Xingguang and Mao, Zhiyuan and Chan, Stanley H.},
  journal={IEEE Transactions on Computational Imaging}, 
  title={Real-Time Dense Field Phase-to-Space Simulation of Imaging Through Atmospheric Turbulence}, 
  year={2022},
  volume={8},
  number={},
  pages={1159-1169},
  doi={10.1109/TCI.2022.3226293}}

@article{gutierrez2006simulation,
  title={Simulation of atmospheric phenomena},
  author={Gutierrez, Diego and Seron, Francisco J and Munoz, Adolfo and Anson, Oscar},
  journal={Computers \& Graphics},
  volume={30},
  number={6},
  pages={994--1010},
  year={2006},
  publisher={Elsevier}
}

@article{bofastsph2016,
author = {Ren, Bo and Yan, Xiao and Yang, Tao and Li, Chen-Feng and Lin, Ming C. and Hu, Shi-Min},
title = {Fast SPH simulation for gaseous fluids},
year = {2016},
issue_date = {April 2016},
publisher = {Springer-Verlag},
address = {Berlin, Heidelberg},
volume = {32},
number = {4},
issn = {0178-2789},
url = {https://doi.org/10.1007/s00371-015-1086-y},
doi = {10.1007/s00371-015-1086-y},
journal = {Vis. Comput.},
month = apr,
pages = {523–534},
numpages = {12}
}

@inproceedings{fraboni2023,
author = {Fraboni, Basile and Chan, Tsz Kin and Vergne, Thibault and Jeziorski, Jakub},
title = {Can You See the Heat? A Null-Scattering Approach for Refractive Volume Rendering},
year = {2023},
isbn = {9798400701436},
publisher = {Association for Computing Machinery},
address = {New York, NY, USA},
url = {https://doi.org/10.1145/3587421.3595427},
doi = {10.1145/3587421.3595427},
booktitle = {ACM SIGGRAPH 2023 Talks},
articleno = {48},
numpages = {2},
location = {Los Angeles, CA, USA},
series = {SIGGRAPH '23}
}

@inproceedings{eikonal2022,
author = {Bemana, Mojtaba and Myszkowski, Karol and Revall Frisvad, Jeppe and Seidel, Hans-Peter and Ritschel, Tobias},
title = {Eikonal Fields for Refractive Novel-View Synthesis},
year = {2022},
isbn = {9781450393379},
publisher = {Association for Computing Machinery},
address = {New York, NY, USA},
url = {https://doi.org/10.1145/3528233.3530706},
doi = {10.1145/3528233.3530706},
booktitle = {ACM SIGGRAPH 2022 Conference Proceedings},
articleno = {39},
numpages = {9},
location = {Vancouver, BC, Canada},
series = {SIGGRAPH '22}
}

@inproceedings{stam96,
author = {Stam, Jos and Langu\'{e}nou, Eric},
title = {Ray tracing in non-constant media},
year = {1996},
isbn = {3211828834},
publisher = {Springer-Verlag},
address = {Berlin, Heidelberg},
booktitle = {Proceedings of the Eurographics Workshop on Rendering Techniques '96},
pages = {225–ff.},
location = {Porto, Portugal}
}

@article{Teh:2022:ANRT,
author = {Arjun Teh and Matthew O'Toole and Ioannis Gkioulekas},
title = {Adjoint Nonlinear Ray Tracing},
journal = {ACM Trans. Graph.},
year = {2022},
publisher = {ACM},
address = {New York, NY, USA},
}

@article{ihrke2007eikonal,
  title={Eikonal rendering: Efficient light transport in refractive objects},
  author={Ihrke, Ivo and Ziegler, Gernot and Tevs, Art and Theobalt, Christian and Magnor, Marcus and Seidel, Hans-Peter},
  journal={ACM Transactions on Graphics (TOG)},
  volume={26},
  number={3},
  pages={59--es},
  year={2007},
  publisher={ACM New York, NY, USA}
}

@article{hardie2017simulation,
  title={Simulation of anisoplanatic imaging through optical turbulence using numerical wave propagation with new validation analysis},
  author={Hardie, Russell C and Power, Jonathan D and LeMaster, Daniel A and Droege, Douglas R and Gladysz, Szymon and Bose-Pillai, Santasri},
  journal={Optical Engineering},
  volume={56},
  number={7},
  pages={071502--071502},
  year={2017},
  publisher={Society of Photo-Optical Instrumentation Engineers}
}

@article{bos2012technique,
  title={Technique for simulating anisoplanatic image formation over long horizontal paths},
  author={Bos, Jeremy P and Roggemann, Michael C},
  journal={Optical Engineering},
  volume={51},
  number={10},
  pages={101704--101704},
  year={2012},
  publisher={Society of Photo-Optical Instrumentation Engineers}
}

@book{schmidt2010numerical,
  title={Numerical simulation of optical wave propagation with examples in MATLAB},
  author={Schmidt, Jason D},
  year={2010},
  publisher={SPIE}
}

@inproceedings{bhattarai2020embedded,
  title={An embedded deep learning system for augmented reality in firefighting applications},
  author={Bhattarai, Manish and Jensen-Curtis, Aura Rose and Mart{\'\i}nez-Ram{\'o}n, Manel},
  booktitle={2020 19th IEEE International Conference on Machine Learning and Applications (ICMLA)},
  pages={1224--1230},
  year={2020},
  organization={IEEE}
}

@article{lanman2013near,
  title={Near-eye light field displays},
  author={Lanman, Douglas and Luebke, David},
  journal={ACM Transactions on Graphics (TOG)},
  volume={32},
  number={6},
  pages={1--10},
  year={2013},
  publisher={ACM New York, NY, USA}
}

@article{daatsim,
  author  = {Saha, Ripon Kumar and Zhang, Yufan and Ye, Jinwei and Jayasuriya, Suren},
  title   = {DAATSim: Depth-Aware Atmospheric Turbulence Simulation for Fast Image Rendering},
  journal = {Computer Graphics Forum},
  year    = {2025}
}

@article{teixeira1998incorporating,
  title={Incorporating turbulence models into the lattice-Boltzmann method},
  author={Teixeira, Christopher M},
  journal={International Journal of Modern Physics C},
  volume={9},
  number={08},
  pages={1159--1175},
  year={1998},
  publisher={World Scientific}
}

@article{jahanshaloo2013review,
  title={A review on the application of the lattice Boltzmann method for turbulent flow simulation},
  author={Jahanshaloo, Leila and Pouryazdanpanah, Emad and Che Sidik, Nor Azwadi},
  journal={Numerical Heat Transfer, Part A: Applications},
  volume={64},
  number={11},
  pages={938--953},
  year={2013},
  publisher={Taylor \& Francis}
}

@inproceedings{saha2024turb,
  title={Turb-seg-res: a segment-then-restore pipeline for dynamic videos with atmospheric turbulence},
  author={Saha, Ripon Kumar and Qin, Dehao and Li, Nianyi and Ye, Jinwei and Jayasuriya, Suren},
  booktitle={Proceedings of the IEEE/CVF Conference on Computer Vision and Pattern Recognition},
  pages={25286--25296},
  year={2024}
}

@inproceedings{zhang2024spatio,
  title={Spatio-temporal turbulence mitigation: a translational perspective},
  author={Zhang, Xingguang and Chimitt, Nicholas and Chi, Yiheng and Mao, Zhiyuan and Chan, Stanley H},
  booktitle={Proceedings of the IEEE/CVF Conference on Computer Vision and Pattern Recognition},
  pages={2889--2899},
  year={2024}
}

@inproceedings{tian2012depth,
  title={Depth from optical turbulence},
  author={Tian, Yuandong and Narasimhan, Srinivasa G and Vannevel, Alan J},
  booktitle={2012 IEEE Conference on Computer Vision and Pattern Recognition},
  pages={246--253},
  year={2012},
  organization={IEEE}
}

@article{chimitt2020simulating,
  title={Simulating anisoplanatic turbulence by sampling intermodal and spatially correlated Zernike coefficients},
  author={Chimitt, Nicholas and Chan, Stanley H},
  journal={Optical Engineering},
  volume={59},
  number={8},
  pages={083101--083101},
  year={2020},
  publisher={Society of Photo-Optical Instrumentation Engineers}
}

@inproceedings{mao2021accelerating,
  title={Accelerating atmospheric turbulence simulation via learned phase-to-space transform},
  author={Mao, Zhiyuan and Chimitt, Nicholas and Chan, Stanley H},
  booktitle={Proceedings of the IEEE/CVF International Conference on Computer Vision},
  pages={14759--14768},
  year={2021}
}

@inproceedings{pbvfs,
author = {Clavet, Simon and Beaudoin, Philippe and Poulin, Pierre},
title = {Particle-based viscoelastic fluid simulation},
year = {2005},
isbn = {1595931988},
publisher = {Association for Computing Machinery},
address = {New York, NY, USA},
url = {https://doi.org/10.1145/1073368.1073400},
doi = {10.1145/1073368.1073400},
booktitle = {Proceedings of the 2005 ACM SIGGRAPH/Eurographics Symposium on Computer Animation},
pages = {219–228},
numpages = {10},
location = {Los Angeles, California},
series = {SCA '05}
}

@techreport{adami2012simulating,
  title={Simulating three-dimensional turbulence with SPH},
  author={Adami, S and Hu, X and Adams, N},
  year={2012},
  institution={Lehrstuhl f{\"u}r Aerodynamik}
}

@article{monaghan2011turbulence,
  title={A turbulence model for smoothed particle hydrodynamics},
  author={Monaghan, Joe J},
  journal={European Journal of Mechanics-B/Fluids},
  volume={30},
  number={4},
  pages={360--370},
  year={2011},
  publisher={Elsevier}
}

@article{violeau2007numerical,
  title={Numerical modelling of complex turbulent free-surface flows with the SPH method: an overview},
  author={Violeau, Damien and Issa, Reza},
  journal={International Journal for Numerical Methods in Fluids},
  volume={53},
  number={2},
  pages={277--304},
  year={2007},
  publisher={Wiley Online Library}
}

@phdthesis{riley2021noise,
  title  = {Noise Function Turbulence Optical Phase Screens and Physics Based Rendering},
  author = {Riley, Joseph T.},
  year   = {2021},
  school = {University of Dayton},
  type   = {Ph.D. dissertation}
}

@article{gladstonedale,
  author  = {Gladstone, J. H. and Dale, T. P.},
  title   = {Researches on the Refraction, Dispersion, and Sensitiveness of Liquids},
  journal = {Philosophical Transactions of the Royal Society of London},
  volume  = {153},
  pages   = {317--343},
  year    = {1863},
  doi     = {10.1098/rstl.1863.0014}
}

@article{lastic2022interactive,
  title={Interactive simulation of plume and pyroclastic volcanic ejections},
  author={Lastic, Maud and Rohmer, Damien and Cordonnier, Guillaume and Jaupart, Claude and Neyret, Fabrice and Cani, Marie-Paule},
  journal={Proceedings of the ACM on Computer Graphics and Interactive Techniques},
  volume={5},
  number={1},
  pages={1--15},
  year={2022},
  publisher={ACM New York, NY, USA}
}

@inproceedings{pretorius2024volcanic,
  title={Volcanic Skies: coupling explosive eruptions with atmospheric simulation to create consistent skyscapes},
  author={Pretorius, P Cilliers and Gain, James and Lastic, Maud and Cordonnier, Guillaume and Chen, Jiong and Rohmer, Damien and Cani, M-P},
  booktitle={Computer Graphics Forum},
  volume={43},
  number={2},
  pages={e15034},
  year={2024},
  organization={Wiley Online Library}
}

@inproceedings{webanck2018procedural,
  title={Procedural cloudscapes},
  author={Webanck, Antoine and Cortial, Yann and Gu{\'e}rin, Eric and Galin, Eric},
  booktitle={Computer Graphics Forum},
  volume={37},
  number={2},
  pages={431--442},
  year={2018},
  organization={Wiley Online Library}
}

@inproceedings{fedkiw2001visual,
  title={Visual simulation of smoke},
  author={Fedkiw, Ronald and Stam, Jos and Jensen, Henrik Wann},
  booktitle={Proceedings of the 28th Annual Conference on Computer Graphics and Interactive Techniques},
  pages={15--22},
  year={2001}
}

@article{monaghan2002sph,
  title={SPH compressible turbulence},
  author={Monaghan, Joseph J},
  journal={Monthly Notices of the Royal Astronomical Society},
  volume={335},
  number={3},
  pages={843--852},
  year={2002},
  publisher={Blackwell Science Ltd}
}

@book{turner1973buoyancy,
  title={Buoyancy effects in fluids},
  author={Turner, John Stewart},
  year={1973},
  publisher={Cambridge University Press}
}

@article{zhang2019finite,
  title={A finite particle method with particle shifting technique for modeling particulate flows with thermal convection},
  author={Zhang, ZL and Walayat, K and Huang, C and Chang, JZ and Liu, MB},
  journal={International Journal of Heat and Mass Transfer},
  volume={128},
  pages={1245--1262},
  year={2019},
  publisher={Elsevier}
}

@article{CLEARY1999227,
title = {Conduction Modelling Using Smoothed Particle Hydrodynamics},
journal = {Journal of Computational Physics},
volume = {148},
number = {1},
pages = {227-264},
year = {1999},
issn = {0021-9991},
doi = {https://doi.org/10.1006/jcph.1998.6118},
url = {https://www.sciencedirect.com/science/article/pii/S0021999198961186},
author = {Paul W Cleary and Joseph J Monaghan}
}

@article{liusph2024,
author = {Liu, Shusen and He, Xiaowei and Guo, Yuzhong and Chang, Yue and Wang, Wencheng},
title = {A Dual-Particle Approach for Incompressible SPH Fluids},
year = {2024},
issue_date = {June 2024},
publisher = {Association for Computing Machinery},
address = {New York, NY, USA},
volume = {43},
number = {3},
issn = {0730-0301},
url = {https://doi.org/10.1145/3649888},
doi = {10.1145/3649888},
journal = {ACM Trans. Graph.},
month = apr,
articleno = {28},
numpages = {18}
}

@inproceedings{ye2024monte,
  title={Monte Carlo Vortical Smoothed Particle Hydrodynamics for Simulating Turbulent Flows},
  author={Ye, Xingyu and Wang, Xiaokun and Xu, Yanrui and Kosinka, Ji{\v{r}}{\'\i} and Telea, Alexandru C and You, Lihua and Zhang, Jian Jun and Chang, Jian},
  booktitle={Computer Graphics Forum},
  volume={43},
  number={2},
  pages={e15024},
  year={2024},
  organization={Wiley Online Library}
}

@article{sph2024,
author = {Probst, Timo and Teschner, Matthias},
title = {Unified Pressure, Surface Tension and Friction for SPH Fluids},
year = {2024},
issue_date = {February 2025},
publisher = {Association for Computing Machinery},
address = {New York, NY, USA},
volume = {44},
number = {1},
issn = {0730-0301},
url = {https://doi.org/10.1145/3708034},
doi = {10.1145/3708034},
journal = {ACM Trans. Graph.},
month = dec,
articleno = {7},
numpages = {28}
}

@article{pbf,
author = {Macklin, Miles and M\"{u}ller, Matthias},
title = {Position based fluids},
year = {2013},
issue_date = {July 2013},
publisher = {Association for Computing Machinery},
address = {New York, NY, USA},
volume = {32},
number = {4},
issn = {0730-0301},
url = {https://doi.org/10.1145/2461912.2461984},
doi = {10.1145/2461912.2461984},
journal = {ACM Trans. Graph.},
month = jul,
articleno = {104},
numpages = {12}
}

@article{lucy1977numerical,
  title={A numerical approach to the testing of the fission hypothesis},
  author={Lucy, Leon B},
  journal={Astronomical Journal, vol. 82, Dec. 1977, p. 1013-1024.},
  volume={82},
  pages={1013--1024},
  year={1977}
}

@article{gingold1977smoothed,
  title={Smoothed particle hydrodynamics: theory and application to non-spherical stars},
  author={Gingold, Robert A and Monaghan, Joseph J},
  journal={Monthly Notices of the Royal Astronomical Society},
  volume={181},
  number={3},
  pages={375--389},
  year={1977},
  publisher={Oxford University Press Oxford, UK}
}

@article{refractiveRTE,
author = {Ament, Marco and Bergmann, Christoph and Weiskopf, Daniel},
title = {Refractive radiative transfer equation},
year = {2014},
issue_date = {March 2014},
publisher = {Association for Computing Machinery},
address = {New York, NY, USA},
volume = {33},
number = {2},
issn = {0730-0301},
url = {https://doi.org/10.1145/2557605},
doi = {10.1145/2557605},
journal = {ACM Trans. Graph.},
month = apr,
articleno = {17},
numpages = {22}
}

@article{seron2005implementation,
  title={Implementation of a method of curved ray tracing for inhomogeneous atmospheres},
  author={Seron, Francisco J and Gutierrez, Diego and Guti{\'e}rrez, Guillermo and Cerezo, Eva},
  journal={Computers \& Graphics},
  volume={29},
  number={1},
  pages={95--108},
  year={2005},
  publisher={Elsevier}
}

@article{berger1990ray,
  title={Ray tracing mirages},
  author={Berger, Marc and Trout, Terry and Levit, Nancy},
  journal={IEEE Computer Graphics and Applications},
  volume={10},
  number={3},
  pages={36--41},
  year={1990},
  publisher={IEEE}
}

@inproceedings{Pfaff2010,
author = {Pfaff, Tobias and Thuerey, Nils and Cohen, Jonathan and Tariq, Sarah and Gross, Markus},
title = {Scalable fluid simulation using anisotropic turbulence particles},
year = {2010},
isbn = {9781450304399},
publisher = {Association for Computing Machinery},
address = {New York, NY, USA},
url = {https://doi.org/10.1145/1866158.1866196},
doi = {10.1145/1866158.1866196},
booktitle = {ACM SIGGRAPH Asia 2010 Papers},
articleno = {174},
numpages = {8},
location = {Seoul, South Korea},
series = {SIGGRAPH ASIA '10}
}

@article{sphheat,
title = {Assessment of Smoothed Particle Hydrodynamics (SPH) models for predicting wall heat transfer rate at complex boundary},
journal = {Engineering Analysis with Boundary Elements},
volume = {111},
pages = {195-205},
year = {2020},
issn = {0955-7997},
doi = {https://doi.org/10.1016/j.enganabound.2019.10.017},
url = {https://www.sciencedirect.com/science/article/pii/S0955799719306320},
author = {K.C. Ng and Y.L. Ng and T.W.H. Sheu and A. Alexiadis}
}

@misc{quickturb,
      title={Turb-Seg-Res: A Segment-then-Restore Pipeline for Dynamic Videos with Atmospheric Turbulence}, 
      author={Ripon Kumar Saha and Dehao Qin and Nianyi Li and Jinwei Ye and Suren Jayasuriya},
      year={2024},
      eprint={2404.13605},
      archivePrefix={arXiv},
      primaryClass={cs.CV},
      url={https://arxiv.org/abs/2404.13605}, 
}

@misc{turbulencesimp2s,
      title={Accelerating Atmospheric Turbulence Simulation via Learned Phase-to-Space Transform}, 
      author={Zhiyuan Mao and Nicholas Chimitt and Stanley H. Chan},
      year={2021},
      eprint={2107.11627},
      archivePrefix={arXiv},
      primaryClass={eess.IV},
      url={https://arxiv.org/abs/2107.11627}, 
}

@article{sissm,
author = {He, Xiaowei and Liu, Shusen and Guo, Yuzhong and Shi, Jian and Qiao, Ying},
title = {A Semi-Implicit SPH Method for Compressible and Incompressible Flows with Improved Convergence},
journal = {Computer Graphics Forum},
volume = {44},
number = {2},
pages = {e70043},
doi = {https://doi.org/10.1111/cgf.70043},
url = {https://onlinelibrary.wiley.com/doi/abs/10.1111/cgf.70043},
year = {2025}
}

@inproceedings{desbrun,
author = {M\"{u}ller, Matthias and Charypar, David and Gross, Markus},
title = {Particle-based fluid simulation for interactive applications},
year = {2003},
isbn = {1581136595},
publisher = {Eurographics Association},
address = {Goslar, DEU},
booktitle = {Proceedings of the 2003 ACM SIGGRAPH/Eurographics Symposium on Computer Animation},
pages = {154–159},
numpages = {6},
location = {San Diego, California},
series = {SCA '03}
}

@inproceedings{desbrun1996smoothed,
  title={Smoothed particles: A new paradigm for animating highly deformable bodies},
  author={Desbrun, Mathieu and Gascuel, Marie-Paule},
  booktitle={Computer Animation and Simulation’96: Proceedings of the Eurographics Workshop in Poitiers, France, August 31--September 1, 1996},
  pages={61--76},
  year={1996},
  organization={Springer}
}

@article{boussinesq1902mise,
  title={Mise en {\'e}quation des ph{\'e}nomenes de convection calorifique et aper{\c{c}}u sur le pouvoir refroidissant des fluides},
  author={Boussinesq, J},
  journal={J. Phys. Theor. Appl.},
  volume={1},
  number={1},
  pages={65--71},
  year={1902}
}

@inproceedings{perlin2002improving,
  title={Improving noise},
  author={Perlin, Ken},
  booktitle={Proceedings of the 29th annual conference on Computer graphics and interactive techniques},
  pages={681--682},
  year={2002}
}


\newpage

\end{document}